\documentclass[lettersize, journal]{IEEEtran}
\usepackage{amsmath,amsfonts,amssymb}
\usepackage{bm}
\usepackage{algorithmic}
\usepackage{algorithm}
\usepackage[caption=false,font=scriptsize,labelfont=sf,textfont=sf]{subfig}
\usepackage{url}
\usepackage{verbatim}
\usepackage{graphicx}
\usepackage{cite}
\usepackage{graphicx}
\usepackage{booktabs}
\usepackage{multirow}
\usepackage{array}
\usepackage{lscape}
\usepackage{adjustbox}
\usepackage{makecell} 
\usepackage{etoolbox}
\usepackage{xcolor}
\usepackage{amsthm}
\usepackage{etoolbox}

\makeatletter
\ifCLASSINFOpdf
\else
\fi
\makeatletter
\patchcmd{\@maketitle}
  {\addvspace{0.5\baselineskip}\egroup}
  {\addvspace{-0.9\baselineskip}\egroup}
  {}
  {}
\makeatother

\begin{document}
\title{Weighted Sum-Rate Enhancement for Flexible Intelligent Metasurface-Assisted Multicell Systems}
\author{Hanwen Hu,~\IEEEmembership{Member,~IEEE}, Jiancheng An,~\IEEEmembership{Senior Member,~IEEE}, Lu Gan,~\IEEEmembership{Member,~IEEE},\\ Hongbin Li,~\IEEEmembership{Fellow,~IEEE}, Naofal Al-Dhahir,~\IEEEmembership{Fellow,~IEEE},\\ George K. Karagiannidis,~\IEEEmembership{Fellow,~IEEE}, and Arumugam Nallanathan,~\IEEEmembership{Fellow,~IEEE}
\thanks{This work was supported by the National Natural Science Foundation of China (NSFC) under Grant 62471096. The work of N. Al-Dhahir was supported by Erik Jonsson Distinguished Professorship at UT-Dallas.  This article was presented in part at
the 2025 IEEE Global Communications Conference (GLOBECOM), Taipei, Taiwan, China, 8-12 December 2025 \cite{11431856}. \textit{(Corresponding Author: Jiancheng An.)}

H. Hu and L. Gan are with the School of Information and Communication
Engineering, University of Electronic Science and Technology of China (UESTC),
Chengdu, Sichuan, 611731, China. L. Gan is also with the Yibin Institute of UESTC, Yibin 644000, China (e-mail: hanwen\_hu\_uestc@outlook.com;
ganlu@uestc.edu.cn). 

J. An is with the School of Electronic Science and Engineering, University of Electronic Science and Technology of China (UESTC), Chengdu, 611731, China (e-mail: jiancheng.an@uestc.edu.cn).

H. Li is with the Department of Electrical and Computer Engineering, Stevens Institute of Technology, Hoboken, NJ 07030 USA (e-mail: hli@stevens.edu). 

N. Al-Dhahir is with the Electrical and Computer Engineering Department,
 University of Texas at Dallas, USA (e-mail: aldhahir@utdallas.edu). 
 
 G. K. Karagiannidis is with the Department of Electrical and Computer Engineering, Aristotle University of Thessaloniki, Greece (e-mail: geokarag@auth.gr). 
 
 A. Nallanathan is with the School of Electronic Engineering and
 Computer Science, Queen Mary University of London, United Kingdom (e-mail: a.nallanathan@qmul.ac.uk).}}
\maketitle
\thispagestyle{empty}
\renewcommand{\abstractname}{\textbf{Abstract}}
\begin{abstract}
\textbf{Flexible intelligent metasurface (FIM) technology has emerged as a promising technology for enhancing wireless communication performance by dynamically reshaping the propagation environment. Compared with conventional rigid reconfigurable intelligent surfaces (RIS), an FIM is composed of multiple electromagnetic (EM) scattering units, each of which can flexibly modify its displacement in the direction normal to the surface, thereby cooperatively morphing the overall surface shape. This additional degree of freedom (DoF) enables improved beamforming and interference mitigation, particularly in complex multicell scenarios. In this paper, an optimization problem for maximizing the weighted sum-rate (WSR) in a multicell multi-user multiple-input single-output (MU-MISO) system assisted by an FIM deployed at the cell boundary is investigated. We jointly optimize the transmit beamforming at the base station (BS), the phase shift matrix, and the FIM surface shape, subject to constraints on the transmit power budget, unit-modulus reflection coefficients, and surface shape morphing range. Due to the non-convex objective function with highly coupled variables, solving the formulated optimization problem is challenging. To tackle this challenge, we propose an efficient alternating optimization framework that leverages the weighted minimum mean square error (WMMSE) method to reformulate the problem and the block coordinate descent (BCD) algorithm to iteratively update the variables. Specifically, the Riemannian conjugate gradient (RCG) algorithm is leveraged to optimize the phase shift matrix, while the projected gradient descent (PGD) method is adopted to optimize the surface shape of the FIM. Additionally, the optimal beamforming vectors are obtained in closed form. Finally, our simulation results demonstrate that the FIM-assisted system achieves an average 33\% improvement in WSR across various scenarios, outperforming conventional RIS schemes.}
\end{abstract}
\IEEEpeerreviewmaketitle

\section{Introduction}
\IEEEPARstart{N}{ext}-generation wireless communication networks are envisioned to deliver substantial enhancements in network capacity and data rate to support the stringent requirements of emerging applications such as augmented reality (AR), embodied intelligence, edge computing, and autonomous driving. To achieve higher spectral efficiency, improved accuracy, and enhanced quality of service (QoS) \cite{ref1,ref2,ref3}, extensive research has been devoted to the development of advanced and innovative wireless communication technologies. Among them, metasurface technology has emerged as a highly promising solution and is widely regarded as a key enabler for sixth-generation (6G) communication and sensing systems \cite{ref4,ref5,ref6,ref7,ref8}. 

Generally speaking, a metasurface is composed of an array of programmable electromagnetic (EM) units that are engineered to control the amplitude, phase, and polarization of incident EM waves\cite{ref9}. By appropriately configuring the phase shifts, a metasurface can reflect and/or refract incoming signals, creating constructive superposition at desired locations, while suppressing unwanted interference. The advantages of deploying metasurfaces in wireless networks are multifaceted. As a passive device without active radio frequency (RF) chains, a metasurface operates with extremely low energy consumption and incurs significantly lower hardware costs compared to traditional systems\cite{ref10}. Moreover, the reconfigurable intelligent surface (RIS) has enabled real-time, dynamic adjustment of phase shifts and amplitudes, thus having significant potential for reconfiguring radio environments\cite{ref11,ref12}. For example, in future underground cities, RIS can facilitate reliable underground signal propagation\cite{ref13}. Additionally, the compact size and lightweight structure of RIS make its deployment highly versatile, allowing installation on building facades, lamp posts, traffic signs, and other existing infrastructures\cite{ref14}. In indoor scenarios, RIS can be mounted on walls to enhance non-line-of-sight (NLoS) WiFi signals when line-of-sight (LoS) paths are obstructed. Another notable advantage of RIS is its compatibility with existing wireless communication technologies\cite{ref15,ref16}. This seamless integration enables RIS to effectively support a wide range of wireless technologies, including multiple-input multiple-output (MIMO) communications, multicell communications, cell-free networks, and millimeter wave (mmWave) communications \cite{ref17,ref18,ref19,ref20,ref21,ref22}.
\begin{table*}[htbp]
 \centering
 \caption{Key Parameters of Existing FIMs}
 \renewcommand{\arraystretch}{1.5} 
 \begin{adjustbox}{max width=\textwidth}
 \begin{tabular}{|c|c|c|c|}
 \hline
 \textbf{Schemes} & \textbf{Ni et al. [38] 2022} & \textbf{Bai et al. [39] 2021} & \textbf{Niu et al. [40] 2022} \\
 \hline
 \textbf{Material} & Silicone \& Liquid metal & Polyimide \& Gold & Hydrogel \& SMP \& Nanocomposite \\
 \hline
 \textbf{Actuation Principle} & Electromagnetic actuation & Electromagnetic actuation& Photomechanical actuation \\
 \hline
 \textbf{Maximum Deformation} & 4 mm& 5.5 mm & 0.35 mm \\
 \hline
 \makecell{\textbf{Relative Maximum Deformation} \\ (Maximum Deformation / Aperture )} & 4 mm / 7.07 mm $\approx$ 0.565 & 5.5 mm / 18 mm $\approx$ 0.3 & 0.35 mm / 0.42 mm $\approx$ 0.825\\
 \hline
 \textbf{Morphing Period} & 30 ms & 10 ms & 500 ms \\
 \hline
 \textbf{Illustration} & \begin{minipage}[b]{0.37\columnwidth}
		\centering
		\raisebox{-.5\height}{\includegraphics[width=\linewidth]{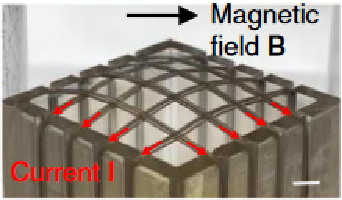}}
	\end{minipage} & \begin{minipage}[b]{0.37\columnwidth}
		\centering
		\raisebox{-.5\height}{\includegraphics[width=\linewidth]{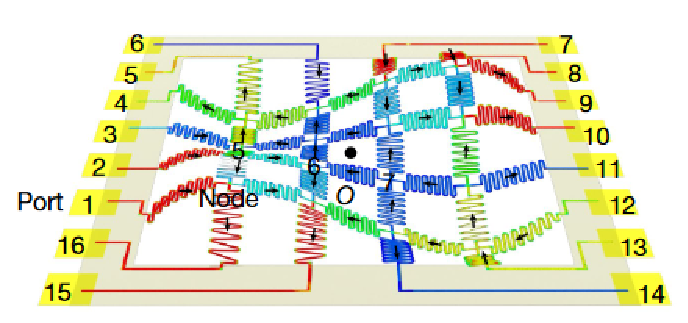}}
	\end{minipage} & \begin{minipage}[b]{0.37\columnwidth}
		\centering
		\raisebox{-.5\height}{\includegraphics[width=\linewidth]{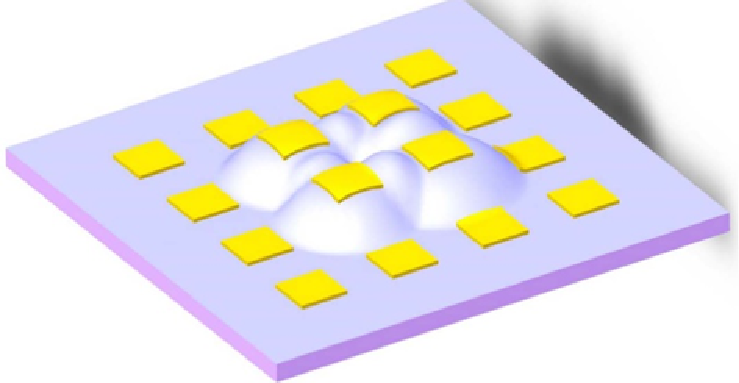}}
	\end{minipage} \\
 \hline
 \end{tabular}
 \end{adjustbox}
\end{table*}

Over the past years, extensive efforts have been dedicated to wireless communication systems aided by RIS, including single-user signal enhancement\cite{ref23}, practical RIS configuration considering discrete phase shifts, RIS-assisted multi-user MIMO downlink communications\cite{ref26,ref27,ref28}, RIS-assisted multicell MIMO scenarios\cite{ref29,ref30}, as well as in RIS-aided secure communication frameworks\cite{ref31,ref32}. To further enhance the potential of RIS and fully exploit its ability to reshape wireless propagation environments, researchers have introduced more advanced forms of intelligent metasurfaces, such as stacked intelligent metasurfaces (SIM)\cite{ref34,ref36,ref37,ref38}. These novel architectures enable more precise control over the EM fields emitted or received through dynamic adjustment of both phase and amplitude. By stacking multiple metasurfaces in a specific configuration, SIM can support machine learning tasks in the wave domain, such as image detection and MIMO precoding\cite{ref33,ref35}, while reducing hardware cost, improving energy efficiency and computational speed.

Nonetheless, the metasurface technologies mentioned above predominantly rely on rigid metamaterials, which may suffer from several inherent limitations. Firstly, their rigid nature limits their ability to conform to complex environments, making deployment challenging in scenarios where flexibility and adaptability are required. Secondly, fixed surface shapes may result in degraded channel quality, particularly under deep fading \cite{ref39}.

The recent discovery and development of flexible metamaterials have paved the way for the realization of flexible intelligent metasurfaces (FIMs), offering unprecedented adaptability and multifunctionality in future wireless communication systems\footnote{\url{https://www.eurekalert.org/multimedia/950133} provides a video demonstrating the real-time morphing ability of an FIM.}. In \cite{ref40}, the authors demonstrated the fabrication of an FIM by embedding nanotube-based mesh structures onto thin silicone layers. To further enhance reconfigurability, gallium-indium liquid metal alloys are injected into the nanotube channels, forming a deformable conductive network. By applying external electromagnetic actuation, the Lorentz force is capable of inducing shape deformation, enabling dynamic reconfiguration of the metasurface's structure and, consequently, its electromagnetic response. In \cite{ref41}, the authors presented a mechanical metasurface that is constructed from a mesh of filamentary metal traces, driven by reprogrammable, distributed Lorentz forces. In \cite{ref42}, a bilayer structure is employed to achieve shape-morphable FIMs through heat or light, which combines a light-absorbing layer with an active layer composed of hydrogels, shape memory polymers (SMPs), and nanocomposites. As FIM technology continues to advance, the spectrum of its applications is broadening accordingly. In \cite{ref43}, the authors provided a comprehensive review of FIMs utilized for cutting-edge tasks, such as image reconstruction of a mannequin and ultrahigh-field magnetic resonance imaging (MRI). To elaborate, Table I shows the main parameters of three typical FIMs.

\begin{table*}[htbp]
 \centering
 \caption{Comparison of Our Contributions to Existing Work}
 \renewcommand{\arraystretch}{1.2}
 \begin{adjustbox}{max width=\textwidth}
 \begin{tabular}{|l|c|c|c|c|c|c|c|c|}
 \hline
 \textbf{Schemes} & \textbf{Metasurface}& \textbf{System Model}& \textbf{$\#$ of Cells} & \textbf{Channel} & \textbf{Optimization Objective} & \textbf{Phase Shift} &\textbf{Deployment}\\
 \hline
 Pan \textit{et al}.\cite{ref21} & RIS & Multi-user MIMO & Multi-cell & Rayleigh & Weighted sum-rate & Continuous &Environment\\
 \hline
 Cui \textit{et al}.\cite{ref34} & RIS & Single-user MISO &Single-cell & Rayleigh &Secrecy rate & Continuous &Environment \\
 \hline
 Wu \textit{et al}.\cite{ref24} & RIS & Multi-user MISO & Single-cell & Rayleigh & Transmit power & Discrete &Environment \\
 \hline
 Guo \textit{et al}.\cite{ref25} & RIS & Multi-user MISO & Single-cell &Rayleigh & Weighted sum-rate & Continuous&Environment\\
 \hline
 An \textit{et al}.\cite{ref44} & FIM & Single-user MIMO & Single-cell & Multipath & Channel capacity & ------ &Transceiver \\
 \hline
 An \textit{et al}.\cite{ref45} & FIM & Multi-user MISO & Single-cell & Multipath & Transmit power& ------ &Transceiver \\
 \hline
 \textbf{Our design} & FIM & Multi-user MISO & Multi-cell& Multipath& Weighted sum-rate & Continuous &Environment \\
 \hline
 \end{tabular}
 \end{adjustbox}
\end{table*}

Compared to traditional rigid metasurfaces, FIMs introduce an additional degree of freedom (DoF) in the spatial domain, enabling three-dimensional (3D) surface shape morphing. In wireless communication systems, this flexibility allows FIMs to dynamically adapt to complex environments and addresses fundamental limitations of conventional rigid metasurfaces, offering improved mitigation of practical multipath fading\cite{ref44,ref45,ref46,ref47,ref48,ref49}. Specifically, in traditional systems, multipath propagation can lead to severe signal fluctuations due to phase misalignment among different propagation paths. However, surface shape morphing enables the constructive superposition of multiple paths at each EM unit. Some existing studies have attempted to deploy FIMs at the transmitter or receiver sides of the system, achieving performance gains beyond conventional MISO/MIMO systems\cite{ref44,ref45,ref46,ref47,ref48,ref49}. However, such fixed deployment locations cannot fully unleash the potential of FIMs. On this basis, deploying the FIM in the environment while simultaneously considering both phase-shifting and surface shape morphing functionalities may provide a better solution for improving the EM environment\cite{ref53}. In particular, environment-deployed FIMs can enhance channels on both sides and can be flexibly installed on curved buildings, wearable fabrics (e.g., invisibility cloaks), or curtains (to improve indoor signal coverage). Despite the promising applications of FIMs and their strong potential to enhance communication performance, practical implementations and algorithmic designs for specific communication scenarios remain largely unexplored. The introduction of shape-morphing capabilities brings additional optimization variables, which are tightly coupled with traditional parameters such as transmit beamforming vectors and phase shift matrix with the unit-modulus constraint.

Against this background, we investigate an FIM-aided multicell multi-user multiple-input single-output (MU-MISO) downlink network where the FIM is strategically deployed at cell boundaries to mitigate inter-cell interference among cell-edge users. Explicitly, we contrast our contributions to the relevant works in Table II. Specifically, the contributions of this paper are summarized as follows.

\begin{itemize}
 \item 
 An FIM is deployed in the environment to enhance multicell MU-MISO downlink communications. Specifically, we formulate an optimization problem aimed at maximizing the weighted sum-rate (WSR) of all users by jointly optimizing the transmit beamforming vectors, the phase shift matrix, and the FIM surface shape, subject to constraints on the transmit power budget, unit-modulus reflection coefficients, and surface shape morphing range. Note that the optimization problem is a multivariable-coupled non-convex optimization. Moreover, due to the multicell structure and the introduction of FIM, a large number of parameters are involved, further complicating the design. To address these challenges, we leverage the weighted minimum mean square error (WMMSE) algorithm to transform the optimization objective into a more tractable formulation\cite{ref50}.
 \item
 Building upon this reformulation, closed-form expressions are derived for both the auxiliary variables and the beamforming vectors. To further handle the coupling between the phase shifts and surface shape, the block coordinate descent (BCD) method is employed to iteratively optimize one variable while keeping the others fixed. For the phase shift matrix optimization, we utilize the Riemannian Conjugate Gradient (RCG) algorithm to obtain a locally optimal solution \cite{ref50,ref51}.

 \item For FIM surface shape optimization, we employ the projected gradient descent (PGD) method. Since the FIM surface shape affects the steering vector associated with each propagation path, the direct gradient calculation seems intractable. To overcome this challenge, we define an FIM kernel to encapsulate the surface shape morphing and multipath interactions, allowing us to leverage matrix vectorization, rank-one decomposition, and the chain rule to derive the gradient in a computationally efficient manner. Furthermore, we approximate the objective function as a quadratic equation via second-order Maclaurin expansion, then derive the optimal step size by minimizing this approximation function.

 \item Our simulation results showcase that the proposed algorithm converges quickly and achieves significant performance improvements compared to conventional RIS-aided communication systems. In particular, FIMs enable the system to take advantage of multipath effects, thereby further improving the network capacity.
\end{itemize}

The rest of this paper is organized as follows. Section~\ref{sec:System Model and Problem Formulation} presents the system model and formulates the WSR maximization problem. In Section~\ref{sec:Block Coordinate Descent Method for FIM-Aided Multicell MU-MISO System}, we employ the BCD algorithm to tackle the joint optimization problem. Section~\ref{sec:Simulation Result} provides extensive numerical results. Finally, conclusions and future research directions are given in Section~\ref{sec:Conclusion}.

Additionally, for clarity, Table III summarizes the main notations used in this paper and their meanings.

\textit{Notations}: Scalars are denoted by italic letters 
(\textit{a, b, ...}). Vectors and matrices are denoted by bold-face lower-case ($\mathbf{a, b}$, ...) and bold-face upper-case letters ($\mathbf{A,B}$, ...). For a vector $\mathbf{a}$, its $i$-th element is denoted by $a_i$. \(\mathbb{E}\{\cdot\}\) denotes the expectation operation. And $j$ = $\sqrt{-1}$ denotes the imaginary unit. The transpose, conjugate transpose are denoted by $(\cdot)^\mathrm{T}$, $(\cdot)^\mathrm{H}$, while $(\cdot)^*$, and $||\cdot ||^2$ denote the complex conjugate, and 2-norm, respectively. $\otimes$ denotes the Kronecker product and $\odot$ denotes the Hadamard product. $\mathbb{C}^{M}$ denotes the set of $M \times 1$ complex vectors. The sets of $x \times y$ complex-valued and real-valued matrices are represented by $\mathbb{C}^{x \times y}$ and $\mathbb{R}^{x \times y}$, respectively. $\operatorname{diag}(\mathbf{x})$ is the diagonal matrix with the vector $\mathbf{x}$ on its diagonal. $\mathcal{CN}(0, \mathbf{A})$ represents the complex-valued Gaussian distribution with mean zero and covariance $\mathbf{A}$.

\begin{table}[t]
\centering
\small 
\caption{List of Major Symbols and Their Meanings}
\label{table:notations}
\begin{tabular}{p{2.3cm} p{5.5cm}} 
\toprule
\textbf{Symbol} & \textbf{Description} \\
\midrule
\multicolumn{2}{l}{\textit{System Parameters}} \\
\midrule
$L, K, M$ & \# of cells, users per cell, BS antennas. \\
$N$ & \# of FIM units ($N = N_y \times N_z$). \\
$P, Q$ & \# of paths in FIM-UE and BS-FIM links. \\
$\lambda, \kappa$ & Carrier wavelength, wavenumber. \\
$P_t, \sigma^2$ & Max BS transmit power, noise power. \\
\midrule
\multicolumn{2}{l}{\textit{Channel Parameters}} \\
\midrule
$\omega_{l,k}$ & User priority weight.  \\
$d_{\text{max}}$ & Morphing range per unit. \\
$\beta_{\hat{l},q}$, $\alpha_{l,k,p}$ & Path loss (BS-FIM, FIM-UE). \\
$\phi^I_{\hat{l},q}$, $\theta^I_{\hat{l},q}$ & Angles of $q$-th BS$ \to$ FIM path. \\
$\phi^O_{l,k,p}$, $\theta^O_{l,k,p}$ & Angles of $p$-th FIM$ \to$ UE path. \\
$\mathbf{h}^r_{l,k}$, $\mathbf{G}_{\hat{l}}$ & Channel: FIM$ \to$ UE, BS$ \to$ FIM. \\
$\bar{\mathbf{h}}_{\hat{l},l,k}$ & Direct channel BS$ \to$ UE. \\
$\mathbf{h}_{\hat{l},l,k}$ & Overall effective channel BS$ \to$ UE. \\
\midrule
\multicolumn{2}{l}{\textit{Optimization Variables}} \\
\midrule
$\mathbf{W}$ & Transmit beamforming matrix. \\
$\boldsymbol{\vartheta}$, $\mathbf{\Phi}$ & FIM phase shift vector and matrix. \\
$\mathbf{u}$, $\mathbf{v}$ & Decoding factors, auxiliary variables. \\
$\mathbf{d}$ & FIM surface shape vector ($|d_n| \leq d_{\text{max}}$).\\
\bottomrule
\end{tabular}
\end{table}

\section{System Model and Problem Formulation}
\label{sec:System Model and Problem Formulation}
\begin{figure}[t] 
\centering %
\includegraphics[width=0.45\textwidth]{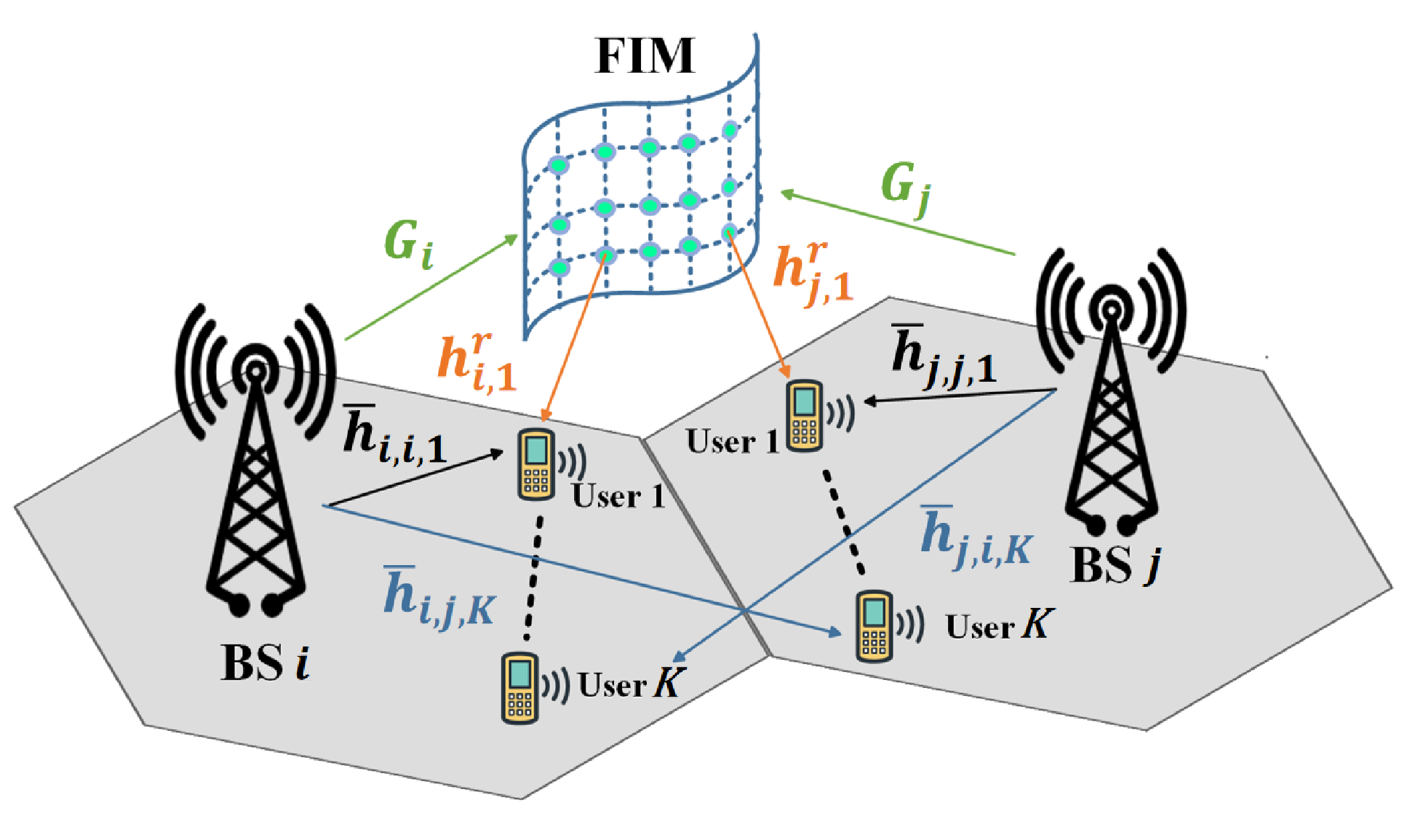}
\caption{An FIM-aided multicell MU-MISO communication system.} 
\label{fig_1}
\vspace{-0.5cm}
\end{figure}
\subsection{System Model}
Fig. 1 illustrates the considered FIM-enhanced multicell downlink MU-MISO communication system. The system is composed of $L$ separate micro cells, each of which contains a base station (BS) equipped with $M$ transmit antennas (TAs), configured as a uniform linear array (ULA) with an inter-element spacing of half a wavelength. Each BS simultaneously serves $K$ single-antenna users (UEs) in its cell. In a multicell communication system, UEs generally experience severely attenuated signals from their serving BS, along with substantial interference from adjacent BSs. To address these issues, we consider an FIM deployed at the cell boundary to assist in the communications, which operates at a carrier wavelength $\lambda$. Specifically, the FIM consisting of $N = N_yN_z$ EM units is situated on the $y$-$z$ plane. $N_y$ and $N_z$ respectively represent the count of flexible EM units arranged parallel to the $y$-axis and $z$-axis.

In sharp contrast to a conventional rigid RIS, the individual EM units of an FIM possess the capability to modify its displacement in the direction normal to the surface, thereby enabling dynamic reconfiguration of the overall surface shape morphology, which is termed as \emph{surface shape morphing}\cite{ref45}. Specifically, the FIM surface shape is represented as $\mathbf{d}=\{d_{1},d_{2},\ldots,d_{N}\}\in\mathbb{R}^{N}$ with $d_n$ representing the displacement of unit $n$. Furthermore, the deformation distance of each unit is restricted to the maximum deformation range of the practical FIM.\cite{ref46}. The deformation capability of the FIM is assumed to be symmetric in both directions of the $x$-axis, and the deformation distance of each element $d_n$ is assumed to satisfy 
 \begin{align}
 |d_n| \leq d_{\text{max}}, \, \forall n = 1, \ldots, N, 
 \label{eq_1}
 \end{align}
 where $d_{\text{max}}$ denotes the maximum deformation distance on one side of the FIM, characterizing the morphing range of reversible deformation\cite{ref46}.
\begin{figure}[t] 
\centering %
\includegraphics[width=0.45\textwidth]{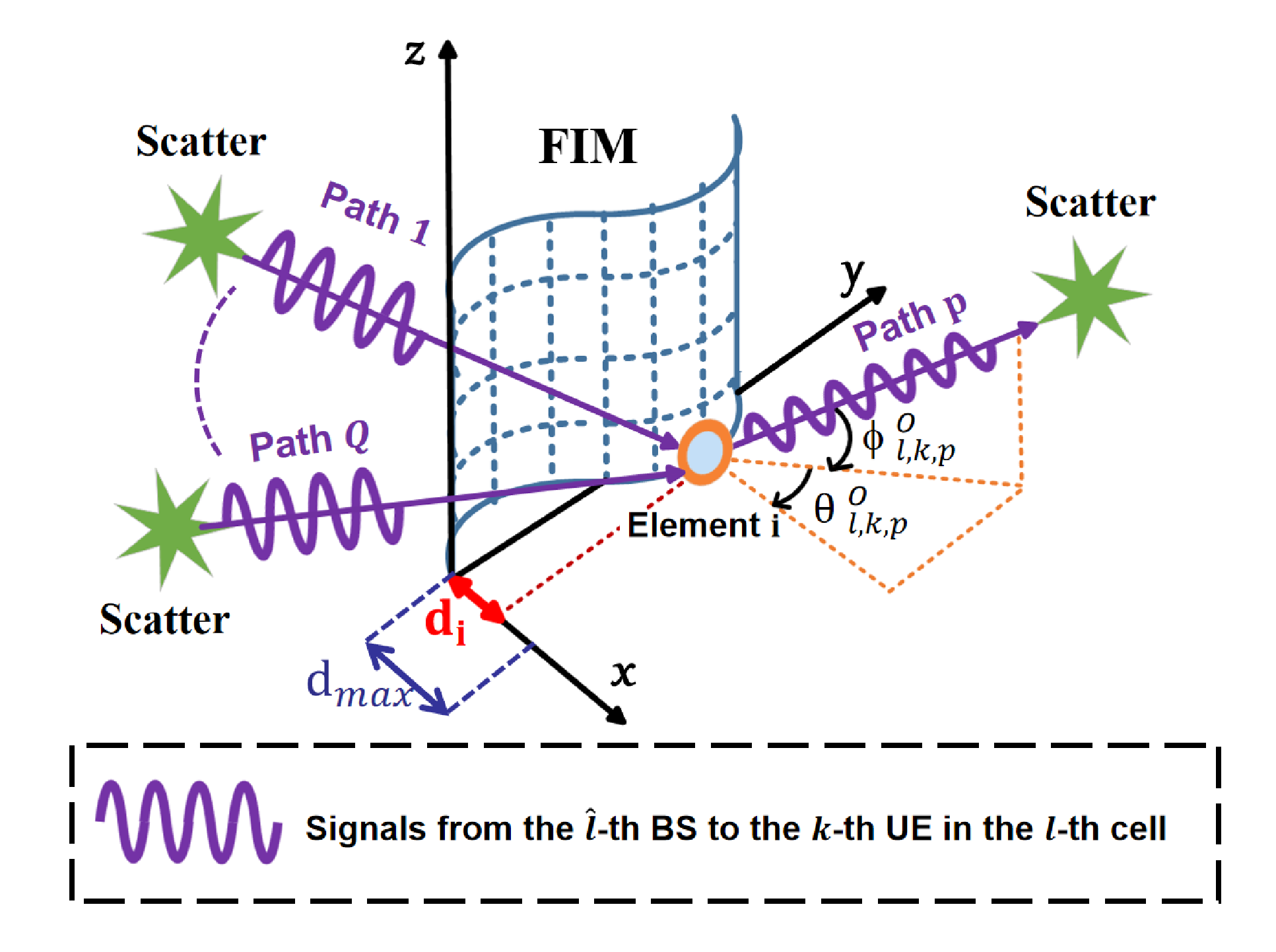}
\caption{An illustration of the key parameters for an FIM-aided system, including the EM waves on both sides with their elevation and azimuth angles, the displacement and morphing range of FIM elements, and the multipath effects due to the presence of scatterers.}
\label{fig_1}
\vspace{-0.5cm}
\end{figure}
\subsection{Multipath Channel Model}
In this paper, we investigate narrow-band multipath channels under block fading conditions, concentrating on a quasi-static block. As mentioned earlier, the channels between the FIM and the BS, as well as those between the FIM and the UEs, are determined by the propagation environment and the FIM surface shape. To simplify the analysis, the system proceeds under the far-field assumption, where the sizes of the FIM and the BS are much smaller than the propagation distance. Therefore, for each propagation path between the BS and the FIM and that between the FIM and the UEs, the channel state information (CSI) remains constant as FIM morphs its surface shape, while only the signal phase varies.

By modeling the unmorphed FIM as a uniform planar array (UPA), the corresponding steering vector $\mathbf{s}_{\text{un}}(\theta, \phi)\in \mathbb{C}^{N}$ is given by
\begin{equation}
\mathbf{s}_{\text{un}}(\theta, \phi) = \mathbf{s}_z( \phi) \otimes\ \mathbf{s}_y(\theta,\phi),
\end{equation}
where $\theta$ and $\phi$ are the azimuth angle and the elevation angle, respectively. $\mathbf{s}_z( \phi)\in \mathbb{C}^{N_y}$ \text{and} $\mathbf{s}_y(\theta,\phi) \in \mathbb{C}^{N_z}$ are represented as
\begin{equation}
\mathbf{s}_z(\phi) = 
\begin{bmatrix}
1,e^{j\pi\sin{\phi}},\ldots,e^{j\pi\sin{\phi}(N_z-1)} 
\end{bmatrix}^\mathrm{T},
\end{equation}
\begin{equation}
\mathbf{s}_y(\theta, \phi) = 
\begin{bmatrix} 
1,e^{j\pi\sin{\theta}\cos{\phi}},\ldots,e^{j\pi\sin{\theta}\cos{\phi}(N_y-1)} 
\end{bmatrix}^\mathrm{T}.
\end{equation}

Furthermore, the surface shape morphing introduces an additional response $\mathbf{s}_{\text{d}}(\theta,\phi,\mathbf{d})$ $\in \mathbb{C}^{N}$, which is given by
\begin{equation}
\mathbf{s}_{\text{d}}(\theta,\phi,\mathbf{d})=e^{j\kappa\cos{\theta}\cos{\phi}\mathbf{d}},
\end{equation}
where $\kappa = 2\pi/\lambda$ is the wavenumber. Therefore, we can derive the overall steering vector of the FIM $\mathbf{s}$($\theta,\phi, \mathbf{d}$) $\in \mathbb{C}^{N}$ as \cite{ref53}
\begin{equation}
 \mathbf{s}(\theta,\phi,\mathbf{d})=
 \mathbf{s}_{\text{un}}(\theta, \phi) \odot \mathbf{s}_{\text{d}}(\theta,\phi,\mathbf{d}).
\end{equation}

For a given transmit angle $\gamma$, the steering vector of the BS $\mathbf{s}_{\text{ula}}(\gamma)\in \mathbb{C}^{M}$ is derived as

\begin{equation}
\mathbf{s}_{\text{ula}}(\gamma) = 
\begin{bmatrix}
1,e^{j\pi\sin{\gamma}},\ldots,e^{j\pi(M-1)\sin{\gamma}} 
\end{bmatrix}^\mathrm{T}.
\end{equation}

To simplify the subsequent calculations, we assume an identical number of paths for the BS-FIM link across different cells, and an identical number of paths from the FIM to UEs in different cells. Let $P$ and $Q$ represent the number of paths from the FIM to the UEs and from the BS to the FIM, respectively. The elevation and azimuth angles of the $p$-th path from the FIM to the $k$-th user in the $l$-th cell are defined as $\phi_{l,k,p}^O$ and $\theta_{l,k,p}^O$, respectively, while the corresponding channel gain is denoted as $\alpha_{l,k,p}$. The transmit angles, the elevation and azimuth angles of the $q$-th path from the BS in the $\hat{l}$-th cell to the FIM are defined as $\gamma_{\hat{l},q}$, $\phi_{\hat{l},q}^I$ and $\theta_{\hat{l},q}^I$, respectively, while the corresponding channel gain is denoted as $\beta_{\hat{l},q}$. The direct channels from the $\hat{l}$-th BS to the $k$-th user in the $l$-th cell is represented as $\mathbf{\bar{h}}_{\hat{l},l,k}$ $\in \mathbb{C}^{1 \times M}$. Additionally, $\mathbf{h}_{l,k}^r$ $\in \mathbb{C}^{1 \times N}$ and $\mathbf{G}_{\hat{l}}$ $\in \mathbb{C}^{N \times M}$ denote the equivalent baseband channels from the FIM to the $k$-th user in the $l$-th cell, and the ones from the $\hat{l}$-th BS to the FIM, respectively. As a result, the narrowband channel $\mathbf{h}_{l,k}^r$ and $\mathbf{G}_{\hat{l}}$ can be written as
 \begin{equation}
 \mathbf{h}_{l,k}^r = \sum_{p=1}^{P} \alpha_{l,k,p} \mathbf{s}^\mathrm{H}(\theta_{l,k,p}^O,\phi_{l,k,p}^O,\mathbf{d})=\sum_{p=1}^{P}\mathbf{a}_{\mathbf{h}_{l,k,p}},
 \end{equation}
\begin{equation}
 \mathbf{G}_{\hat{l}} = \sum_{q=1}^{Q} \beta_{\hat{l},q} \mathbf{s}(\theta_{\hat{l},q}^I,\phi_{\hat{l},q}^I,\mathbf{d})\mathbf{s}_{\text{ula}}^\mathrm{H}(\gamma_{\hat{l},q})=\sum_{q=1}^{Q}\beta_{\hat{l},q}\mathbf{a_g}_{\hat{l},q}\mathbf{a_{bs}}_{\hat{l},q}^\mathrm{H}.
\end{equation}

As signals interact with the FIM, each FIM unit imparts a phase shift $\varphi_n$ to the incident multipath signals and the modified waves are coherently combined for transmission to the UEs. Let $\vartheta_n \overset{\triangle}{=} e^{j\varphi_n}$ and  define a diagonal matrix $\mathbf{\Phi} = \operatorname{diag}(\boldsymbol{\vartheta})$ as the phase shift matrix of the FIM, where $\boldsymbol{\vartheta}\overset{\triangle}{=}\left[ \vartheta_1, \ldots,\vartheta_n ,\ldots, \vartheta_N \right]^\mathrm{H}$. The equivalent channel $\mathbf{h}_{\hat{l},l,k}$ $\in \mathbb{C}^{1\times M}$ spanning from the $\hat{l}$-th BS to the $k$-th user in the $l$-th cell is the sum of the direct channel and the cascaded channel reflected by the FIM, which is given by
\begin{equation}
 \mathbf{h}_{\hat{l},l,k} = \mathbf{\bar{h}}_{\hat{l},l,k}+ \mathbf{h}_{l,k}^r\mathbf{\Phi}\mathbf{G}_{\hat{l}}.
\end{equation}

Furthermore, we assume that the CSI of all channels is perfectly known at the BSs, which allows us to characterize an upper bound on the performance in real-world scenarios.

\subsection{Signal Model}
We assume that spatial division multiple access strategy is utilized to enable concurrent multi-user communications within identical time-frequency resource blocks by spatially multiplexing distinct data streams. The BS employs an active transmit beamforming strategy to send signals from its $M$ antennas. Let $s_{l,k}$ and $\mathbf{w}_{l,k}$ $\in \mathbb{C}^{M \times 1}$ denote the transmitted data and transmit beamforming vector for the $k$-th UE in the $l$-th cell, respectively. The data symbol $s_{l,k}$ satisfies $\mathbb{E} \left[ s_{l,k} s_{l,k}^\mathrm{H} \right] = 1$ and $\mathbb{E} \left[ s_{l,k} s_{i,j}^\mathrm{H} \right] = 0$, for $\{l, k\} \neq \{i, j\}$.
Hence, the signal transmitted by the $l$-th BS $\mathbf{x}_l$ $\in \mathbb{C}^{M \times 1}$ is given by
\begin{equation}
 \mathbf{x}_l = \sum_{k=1}^K \mathbf{w}_{l,k}s_{l,k}.
 \label{eq_10}
\end{equation}

Hence, the signal $y_{l,k}$ received by the $k$-th user in the $l$-th cell is given by
\begin{align}
 y_{l,k} &= \sum_{\hat{l}=1}^L\mathbf{h}_{\hat{l},l,k}\mathbf{x}_{\hat{l}} + {n}_{l,k}, \label{eq_11}
\end{align}
where $n_{l,k}\sim \mathcal{CN}(0, \sigma_{l,k}^2)$ represents the additive white Gaussian noise (AWGN) at the $k$-th user in the $l$-th cell, with $\sigma_{l,k}^2$ being the corresponding noise power.

By substituting \eqref{eq_10} into \eqref{eq_11}, we can rewrite $y_{l,k}$ as
\begin{align}
 y_{l,k} &= \sum_{\hat{l}=1}^L\mathbf{h}_{\hat{l},l,k}(\sum_{\hat{k}=1}^K \mathbf{w}_{\hat{l},\hat{k}}s_{\hat{l},\hat{k}})+ n_{l,k} \notag \\
 &= \mathbf{h}_{l,l,k} \mathbf{w}_{l,k} s_{l,k} 
 +  \underbrace{\sum_{\hat{l}=1, \hat{l} \neq l}^{L} \sum_{\hat{k}=1}^{K} \mathbf{h}_{\hat{l},l,k} \mathbf{w}_{\hat{l},\hat{k}}s_{\hat{l},\hat{k}}}_{\text{Inter-cell interference}}\notag \\
 &+\underbrace{\sum_{\hat{k}=1, \hat{k} \neq k}^{K} \mathbf{h}_{l,l,k} \mathbf{w}_{l,\hat{k}} s_{l,\hat{k}}}_{\text{Intra-cell interference}} 
 + n_{l,k}.
\end{align}

Then, the achievable data rate of the $k$-th user in the $l$-th cell $R_{l,k}(\mathbf{W}, \mathbf{\Phi},\mathbf{d})$ can be derived as 
\begin{align}
 R_{l,k}(\mathbf{W}, \mathbf{\Phi},\mathbf{d}) = \log|1 + \mathbf{h}_{l,l,k} \mathbf{w}_{l,k} \mathbf{w}_{l,k}^\mathrm{H} \mathbf{h}_{l,l,k}^\mathrm{H}\notag\\
 \times (J_{l,k}-\mathbf{h}_{l,l,k} \mathbf{w}_{l,k} \mathbf{w}_{l,k}^\mathrm{H} \mathbf{h}_{l,l,k}^\mathrm{H} )^{-1} |,
\end{align}
where we define the beamforming matrix $\mathbf{W}\overset{\triangle}{=}\left[ \mathbf{w}_{l,k}, \, \forall l, k \right]$, and the received signal covariance $J_{l,k}$ is given as
\begin{align}
 J_{l,k} &= \mathbb{E} \left[ y_{l,k}y_{l,k}^* \right] = \mathbf{h}_{l,l,k} \mathbf{w}_{l,k} \mathbf{w}_{l,k}^\mathrm{H} \mathbf{h}_{l,l,k}^\mathrm{H}\notag\\
 & + \sum_{\hat{l}=1,\hat{l} \neq l}^{L} \sum_{\hat{k}=1}^{K} \mathbf{h}_{\hat{l},l,k} \mathbf{w}_{\hat{l},\hat{k}}\mathbf{w}_{\hat{l},\hat{k}}^\mathrm{H}\mathbf{h}_{\hat{l},l,k} ^\mathrm{H} \notag \\
 & + \sum_{\hat{k}=1, \hat{k} \neq k}^{K} \mathbf{h}_{l,l,k} \mathbf{w}_{l,\hat{k}}\mathbf{w}_{l,\hat{k}}^\mathrm{H}\mathbf{h}_{l,l,k}^\mathrm{H}
 + \sigma_{l,k}^2.
\end{align}

\subsection{Problem Formulation}
This paper focuses on the optimization of maximizing the WSR of all UEs by jointly optimizing the phase shift of the FIM $\mathbf{\Phi}$, the transmit beamforming $\mathbf{W}$ and the FIM surface shape $\mathbf{d}$, subject to constraints on total transmit power, unit-modulus phase shift and morphing range. Accordingly, the problem is formulated as
\begin{align}
 &\max_{\mathbf{W}, \mathbf{\Phi},\mathbf{d}} \quad \sum_{l=1}^{L} \sum_{k=1}^{K} \omega_{l,k} R_{l,k}(\mathbf{W}, \mathbf{\Phi},\mathbf{d}) \tag{16a} \label{eq_15a} \\
 &\quad\text{s.t.} \quad 
 \sum_{k=1}^{K} \| \mathbf{w}_{l,k} \|_F^2 \leq P_{t,l}, \quad\forall l = 1, \ldots, L, \tag{16b} \\
 & \quad \quad \quad|\vartheta_n|=1, \quad\forall n = 1. \ldots, N, \tag{16c} \\
 & \quad \quad \quad|d_n| \leq d_{\text{max}}, \quad\forall n = 1, \ldots, N, \tag{16d}
\end{align}
where $\omega_{l,k}$ is used to represent the priority of user $k$ in the $l$-th cell, and $P_{t,l}$ represents the maximum transmit power of the $l$-th BS. Due to the non-convex objective function \eqref{eq_15a} with highly coupled variables, the optimization problem is difficult to solve. To facilitate analytical tractability and reduce computational complexity, we employ the WMMSE algorithm to recast the objective function \eqref{eq_15a}. 

By introducing a linear decoding vector $\mathbf{u} \overset{\triangle}{=} \{u_{l,k}, \forall l, k\}$, the estimated signal at the $k$-th user in the $l$-th cell is given by
\setcounter{equation}{16}
\begin{equation}
 \hat{s}_{l,k} = u^*_{l,k}y_{l,k},
\end{equation}
where $u_{l,k}$ is the decoding factor for corresponding user. Since the received signal $y_{l,k}$ and noise $n_{l,k}$ are mutually independent, the expected mean-square error (MSE) factor can be derived as
\begin{align}
e_{l,k} &= \mathbb{E}_{s_{l,k},n_{l,k}} \left[ (\hat{s}_{l,k} - s_{l,k}) (\hat{s}_{l,k} - s_{l,k})^* \right] \label{eq:first} \\
&= (u^*_{l,k} \mathbf{h}_{l,l,k}\mathbf{w}_{l,k} - 1) (u^*_{l,k} \mathbf{h}_{l,l,k}\mathbf{w}_{l,k} - 1)^* \notag \\
&+ \sum_{\hat{k}=1,\hat{k}\neq k}^K \left|u_{l,k}\right|^2 \mathbf{h}_{l,l,k} \mathbf{w}_{l,\hat{k}} \mathbf{w}^\mathrm{H}_{l,\hat{k}} \mathbf{h}^\mathrm{H}_{l,l,k} \notag \\
&+ \sum_{\hat{l}=1,\hat{l}\neq l}^L \sum_{\hat{k}=1}^K \left|u_{l,k}\right|^2\mathbf{h}_{\hat{l},l,k} \mathbf{w}_{\hat{l},\hat{k}} \mathbf{w}^\mathrm{H}_{\hat{l},\hat{k}} \mathbf{h}^\mathrm{H}_{\hat{l},l,k} \notag \\
&+ \left|u_{l,k}\right|^2\sigma_{l,k}^2 , \quad \forall l,k.
\label{eq_18}\end{align}

Then, we introduce a group of auxiliary parameters $\mathbf{v} \overset{\triangle}{=} \{v_{l,k}, \forall l, k\}$ and establish a more tractable WMMSE problem which is equal to \eqref{eq_15a} as follows\cite{ref50}
\begin{align}
 &\min_{\mathbf{W}, \mathbf{\Phi},\mathbf{d},\mathbf{v,u}} \quad \sum_{l=1}^{L} \sum_{k=1}^{K} \omega_{l,k} (v_{l,k}e_{l,k}-\log |v_{l,k}|)\tag{20a}\label{eq_19a} \\
 &\quad \quad \text{s.t.} \quad \quad \quad \text{(16b), (16c), (16d)}. \tag{20b}
\end{align}

\noindent\textbf{Remark 1:} The equivalence between \eqref{eq_15a} and \eqref{eq_19a} can be proved by deriving the optimal $v_{l,k},u_{l,k}, \forall l,k$, in \eqref{eq_19a} and substituting them back into \eqref{eq_15a}\cite{ref50}.

\section{Block Coordinate Descent Method for FIM-Aided Multicell MU-MISO System}
\label{sec:Block Coordinate Descent Method for FIM-Aided Multicell MU-MISO System}
In this section, we propose a BCD method to solve the challenging WSR optimization problem. The core idea is to partition the variables into four blocks: (i) the decoding factor $\mathbf{u}$ and auxiliary variable $\mathbf{v}$; (ii) the transmit beamforming matrix $\mathbf{W}$; (iii) the FIM phase shift matrix $\mathbf{\Phi}$; and (iv) the FIM surface shape vector $\mathbf{d}$. These blocks are then optimized alternately in each iteration. The detailed procedures for each block are presented in the following subsections.

\subsection{Optimizing the Decoding and the Auxiliary Vectors $\{\mathbf{u},\mathbf{v}\}$}
In this subsection, the optimization of the newly introduced decoding and auxiliary vectors will be discussed. Specifically, for given $\mathbf{d}$, $\mathbf{\Phi}$, $\mathbf{W}$ and $\mathbf{v}$, the optimal decoding factor $\hat{u}_{l,k}$ is given as \cite{ref21,ref50}
\setcounter{equation}{20}
\begin{equation}
 \hat{u}_{l,k}= J_{l,k}^{-1}\mathbf{h}_{l,l,k}\mathbf{w}_{l,k}.
\end{equation}

Similarly, for given $\mathbf{d}$, $\mathbf{\Phi}$, $\mathbf{W}$ and $\mathbf{u}$, the optimal auxiliary parameter $\hat{v}_{l,k}$ can be obtained as \cite{ref21,ref50}
\begin{equation}
 \hat{v}_{l,k} = \hat{e}_{l,k}^{-1},
\end{equation}
where the optimal MSE factor $\hat{e}_{l,k}$ is obtained by plugging the expression of $\hat{u}_{l,k}$ into \eqref{eq_18}, yielding
\begin{align}
 \hat{e}_{l,k} &= 1-\hat{u}_{l,k}^*\mathbf{h}_{l,l,k}\mathbf{w}_{l,k} \notag\\
 &=1 - J_{l,k}^{-1}\mathbf{w}_{l,k}^\mathrm{H}\mathbf{h}_{l,l,k}^\mathrm{H} \mathbf{h}_{l,l,k}\mathbf{w}_{l,k}.
\end{align}

\subsection{Optimizing the Transmit Beamforming $\mathbf{W}$}
In this subsection, we focus on the optimal design of the beamforming $\mathbf{W}$. According to (20a) and (23), it is clear that the update over $\mathbf{W}$ can be decoupled across different BSs. Thus we have the following optimization problem by substituting $\hat{e}_{l,k}$ into \eqref{eq_19a}:
\begin{align}
 &\max_{\mathbf{w}_{l,k}, \forall k}\quad \sum_{k=1}^{K} \left[2\omega_{l,k} v_{l,k} \operatorname{Re}\left(u_{l,k}^* \mathbf{h}_{l,l,k} \mathbf{w}_{l,k}\right)-\mathbf{w}_{l,k}^\mathrm{H} \mathbf{S}_l \mathbf{w}_{l,k}\right]\tag{24a} \\
 &\quad\quad\text{s.t.} \quad 
 \sum_{k=1}^{K} \| \mathbf{w}_{l,k} \|_F^2 \leq P_{t,l}, \quad\forall l = 1, \ldots, L,\tag{24b} 
\end{align} where 
$\mathbf{S}_l = \sum_{\smash{\hat{l}}=1}^{L} \sum_{\hat{k}=1}^{K} \omega_{\hat{l},\hat{k}} v_{\hat{l},\hat{k}} |u_{\hat{l},\hat{k}}|^2\mathbf{h}_{l,\hat{l},\hat{k}}^\mathrm{H}\mathbf{h}_{l,\hat{l},\hat{k}}\in \mathbb{C}^{M \times M}$. Detailed transformations of (24) is referred to Appendix. The above problem is proved to be a convex quadratic optimization problem and admits a closed-form solution by applying the method of Lagrange multipliers, which is derived as \cite{ref50,ref25}
\setcounter{equation}{24}
\begin{equation}
\mathbf{w}_{l,k}(\mu_l) = \omega_{l,k} u_{l,k} v_{l,k}\left( \mathbf{S}_l + \mu_l \mathbf{I}_M \right)^{-1} \mathbf{h}_{l,l,k}^\mathrm{H},
\end{equation}
where $\mu_l\geq0$ is the dual variable related to the power constraint in (24b). Specifically, when $\mathbf{S}_l$ is invertible and $\sum_{k=1}^{K} \| \mathbf{w}_{l,k}(0) \|_F^2 \leq P_{t,l}$, then $\mathbf{w}_{l,k}^*=\mathbf{w}_{l,k}(0)$. Otherwise, the optimal dual variable $\mu_l^*$ must satisfy 
\begin{equation}
 \sum_{k=1}^{K} \| \mathbf{w}_{l,k}(\mu_l^*) \|_F^2 = P_{t,l}\label{eq_25}.
\end{equation}
Since the left-hand-side of \eqref{eq_25} is a monotonic decreasing function with respect to $\mu_l\geq 0$. Hence, the optimal solution $\mu_l^*$ can be effectively obtained via one dimensional search methods, yielding the corresponding optimal beamforming vector $\mathbf{w}_{l,k}^*=\mathbf{w}_{l,k}(\mu_l^*)$.

\subsection{Optimizing the Phase Shift Matrix $\mathbf{\Phi}$}
\label{sec:phase shift}
This subsection details the optimization of the phase shift matrix $\mathbf{\Phi}$. Specifically, according to the Appendix, given $\mathbf{u}$, $\mathbf{v}$, $\mathbf{W}$ and $\mathbf{d}$ and dropping irrelevant constants, we proceed to reformulate the phase shift optimization problem as
\begin{align}
 &\max_{\mathbf{\Phi}} \ \sum_{l=1}^{L} \sum_{k=1}^{K} 2\omega_{l,k} v_{l,k} \operatorname{Re}\left(u_{l,k}^* \mathbf{h}_{l,l,k} \mathbf{w}_{l,k}\right) \notag \\
 & - \sum_{\smash{\hat{l}}=1}^{L} \sum_{l=1}^{L}\sum_{k=1}^{K} \omega_{l,k} v_{l,k} \left|u_{l,k}\right|^2 \mathbf{h}_{\hat{l},l,k} \mathbf{W}_{\hat{l}}\mathbf{h}_{\hat{l},l,k}^\mathrm{H}\label{eq_26a}\tag{27a} \notag \\
 &\quad\text{s.t.} \quad 0\leq \varphi_n \leq 2\pi, \quad\forall n = 1. \ldots, N,\tag{27b}
\end{align}
where $\mathbf{W}_{\hat{l}}=\sum_{\hat{k}=1}^{K}\mathbf{w}_{\hat{l},\hat{k}} \mathbf{w}_{\hat{l},\hat{k}}^\mathrm{H}$.

To make the above expression more tractable, we define $\mathbf{H}_{\hat{l},l,k}\overset{\triangle}{=}\operatorname{diag}(\mathbf{h}_{l,k}^r)\mathbf{G}_{\hat{l}}$, and then the equivalent channel $\mathbf{h}_{\hat{l},l,k}$ can be represented as
\setcounter{equation}{27}
\begin{equation}
 \mathbf{h}_{\hat{l},l,k} = \mathbf{\bar{h}}_{\hat{l},l,k}+ \boldsymbol{\vartheta}^\mathrm{H}\mathbf{H}_{\hat{l},l,k}.
 \label{eq_27}
\end{equation}

By plugging \eqref{eq_27} into \eqref{eq_26a}, we have 
\begin{align}
 \mathbf{h}_{\hat{l},l,k} \mathbf{W}_{\hat{l}} \mathbf{h}_{\hat{l},l,k}^\mathrm{H}&= \boldsymbol{\vartheta}^\mathrm{H}\mathbf{H}_{\hat{l},l,k} \mathbf{W}_{\hat{l}} \mathbf{H}_{\hat{l},l,k}^\mathrm{H} \boldsymbol{\vartheta}\notag+\mathbf{\bar{h}}_{\hat{l},l,k}\mathbf{W}_{\hat{l}} \mathbf{H}_{\hat{l},l,k}^\mathrm{H} \boldsymbol{\vartheta} \\
 &+ \boldsymbol{\vartheta}^\mathrm{H}\mathbf{H}_{\hat{l},l,k} \mathbf{W}_{\hat{l}} \mathbf{\bar{h}^\mathrm{H}}_{\hat{l},l,k}+\mathbf{\bar{h}}_{\hat{l},l,k}\mathbf{W}_{\hat{l}} \mathbf{\bar{h}^\mathrm{H}}_{\hat{l},l,k},\\
 \mathbf{h}_{l,l,k} \mathbf{w}_{l,k} &= \mathbf{\bar{h}^\mathrm{H}}_{l,l,k} \mathbf{w}_{l,k}+ \boldsymbol{\vartheta}^\mathrm{H}\mathbf{H}_{l,l,k} \mathbf{w}_{l,k}.
\end{align}
\begin{figure*}[t!]
\vspace*{-1cm} % 使用vspace*版本
\begin{align}
&\min_{\mathbf{d}} \sum_{\smash{\hat{l}}=1}^{L} \sum_{l=1}^{L} \sum_{\smash{\hat{k}}=1}^{K}\sum_{k=1}^{K}\omega_{l,k} v_{l,k}\left|u_{l,k}\right|^2\left[\boldsymbol{\vartheta}^\mathrm{H}\mathbf{H}_{\hat{l},l,k}\mathbf{w}_{\hat{l},\hat{k}}\mathbf{w}_{\hat{l},\hat{k}}^\mathrm{H}\mathbf{H}_{\hat{l},l,k}^\mathrm{H}\boldsymbol{\vartheta}+2\operatorname{Re}\left(\boldsymbol{\vartheta}^\mathrm{H}\mathbf{H}_{\hat{l},l,k}\mathbf{w}_{\hat{l},\hat{k}}\mathbf{w}_{\hat{l},\hat{k}}^\mathrm{H}\mathbf{\bar{h}^\mathrm{H}}_{\hat{l},l,k}\right)\right]\notag\\
&\quad- \sum_{l=1}^{L} \sum_{k=1}^{K}\operatorname{Tr}\left[ 2\omega_{l,k}\operatorname{Re}\left(v_{l,k}u_{l,k}^*\mathbf{w}_{l,k}\boldsymbol{\vartheta}^\mathrm{H}\mathbf{H}_{l,l,k}\right)\right]\tag{39a}\label{eq_37a}\\
& \quad\text{s.t.} \quad|d_n| \leq d_{\text{max}}, \quad\forall n = 1, \ldots, N.\tag{39b}
\end{align}
\hrulefill
\end{figure*}
To simplify the above equations, we merge the factors unrelated to the phase shift $\boldsymbol{\vartheta}$, while ignoring constants independent of $\boldsymbol{\vartheta}$, resulting in the following problem equivalent to \eqref{eq_26a} as
\begin{align}
 &\min_{\boldsymbol{\vartheta}} \quad f\left(\boldsymbol{\vartheta}\right)=\boldsymbol{\vartheta}^\mathrm{H}\mathbf{Z}\boldsymbol{\vartheta}+2\operatorname{Re}\left(\boldsymbol{\vartheta}^\mathrm{H}\boldsymbol{\nu}\right)\tag{31a}\label{eq_30a}\\
 &\quad\text{s.t.} \quad \left|\vartheta_n\right|=1, \quad\forall n = 1. \ldots, N,\label{eq_30b}\tag{31b}
\end{align}
where $\mathbf{Z}$ and $\boldsymbol{\nu}$ can be represented as 
\setcounter{equation}{31}
\begin{equation}
 \mathbf{Z}= \sum_{\smash{\hat{l}}=1}^{L} \sum_{l=1}^{L}\sum_{k=1}^{K} \omega_{l,k}v_{l,k}\left|u_{l,k}\right|^2\mathbf{H}_{\hat{l},l,k}\mathbf{W}_{\hat{l}} \mathbf{H}_{\hat{l},l,k}^\mathrm{H},
\end{equation}
\begin{align}
 \boldsymbol{\nu}&=\sum_{\smash{\hat{l}}=1}^{L} \sum_{l=1}^{L} \sum_{k=1}^{K}
\omega_{l,k}v_{l,k}\left|u_{l,k}\right|^2\textbf{}\mathbf{H}_{\hat{l},l,k} \mathbf{W}_{\hat{l}} \mathbf{\bar{h}^\mathrm{H}}_{\hat{l},l,k} \notag\\
 &- \sum_{l=1}^{L} \sum_{k=1}^{K}\omega_{l,k}v_{l,k}u_{l,k}^*\mathbf{H}_{l,l,k} \mathbf{w}_{l,k}.
\end{align}

It is evident that the unit-modulus constraints \eqref{eq_30b} makes problem (31) inherently non-convex. To tackle this, we propose an efficient Riemannian conjugate gradient (RCG) algorithm since the unit-modulus constraints in (31b) form a complex circle manifold $\mathcal{M} = \{ \boldsymbol{\vartheta} \in \mathbb{C}^N : |\mathbf{\vartheta}_1| = \cdots = |\mathbf{\vartheta}_N|=1 \}$. The main steps of RCG in each iteration are summarized as follows \cite{ref21}:
\subsubsection{Riemannian Gradients}Geometrically, the Riemannian gradient $\operatorname{grad} f$ on the manifold $\mathcal{M}$ is given by projecting the Euclidean gradient $\nabla f$ onto $\mathcal{M}$ along the orthogonal direction. Specifically, at the $i$-th iteration, the Riemannian gradient $\operatorname{grad}_if$ is
\begin{equation}
 \operatorname{grad}_if = \nabla_i f - \operatorname{Re}\left(\nabla_i f \odot \boldsymbol{\vartheta}_i^*\right)\odot\boldsymbol{\vartheta}_i,
\end{equation}
where the Euclidean gradient $\nabla_i f$ is given by
\begin{equation}
 \nabla_i f = 2\mathbf{Z}\boldsymbol{\vartheta}_i+\boldsymbol{\nu}.
\end{equation}
\subsubsection{Search Direction} The update rule of the search direction for
the RCG method on manifolds can be given by
\begin{equation}
 \boldsymbol{\eta}_{i} = -\operatorname{grad}_{i} f + \tau_1 \mathcal{T}_{i-1\to i}(\boldsymbol{\eta}_{i-1}),
\end{equation}
where the function $\mathcal{T}_{i-1\to i}(\cdot)$ for transporting the tangent vector from the previous search direction to the current tangent space is defined as
\begin{equation}
 \mathcal{T}_{i-1\to i}(\boldsymbol{\eta})=\boldsymbol{\eta}_{i-1}-\operatorname{Re}\left(\boldsymbol{\eta}_{i-1} \odot \boldsymbol{\vartheta}_{i}^*\right)\odot\boldsymbol{\vartheta}_{i},
\end{equation}
and $\tau_1$ is the conjugate gradient Polak-Ribiere parameter.
\subsubsection{Retraction Operator}Project the updated phase shift vector onto the complex circle via vector normalization as 

\begin{equation}
 \boldsymbol{\vartheta}_{i+1} = \frac{(\boldsymbol{\vartheta}_i + \tau_2 \boldsymbol{\eta}_{i})}{|(\boldsymbol{\vartheta}_i + \tau_2 \boldsymbol{\eta}_{i})|},
\end{equation}
where $\tau_2$ is the Armijo step size.

The details of the RCG algorithm are presented in Algorithm 1. In addition, the worst-case total computational complexity is $\mathcal{O}(L_{RCG}N^{1.5})$ according to \cite{ref21}, where $L_{RCG}$ represents the number of iterations for the RCG algorithm. 
\begin{algorithm}[t]
\renewcommand{\algorithmicrequire}{\textbf{Input:}}
\renewcommand{\algorithmicensure}{\textbf{Output:}}
\renewcommand{\algorithmicrepeat}{\textbf{Repeat}}
\renewcommand{\algorithmicuntil}{\textbf{Until}}
\caption{Riemannian Conjugate Gradient (RCG) on the Complex Circle Manifold}
\begin{algorithmic}[1]
\STATE Initialize the iteration number $i$ = 0 and the accuracy $\epsilon_1$. 
\STATE Construct a feasible solution $\boldsymbol{\vartheta}_0$. \STATE Calculate the value of the objective function in \eqref{eq_30a} as $f(\boldsymbol{\vartheta}_0$).
\STATE Calculate the initial search direction $\boldsymbol{\eta}_0 = -\operatorname{grad}_0 f$.
\REPEAT
 \STATE Set the Armijo backtracking line search step size $\tau_2$.
 \STATE Calculate the next feasible solution $\boldsymbol{\vartheta}_{i+1}$ in (38). 
 \STATE Determine the Riemannian gradient $\operatorname{grad}_{i+1} f$ in (34).
 \STATE Choose the Polak-Ribiere parameter $\tau_1$.
 \STATE Update the search direction $\boldsymbol{\eta}_{i+1}$ by (36).
 \STATE Calculate the value of the objective function as $f(\boldsymbol{\vartheta}_{i+1}$).
 \STATE Set $i+1\to i$.
\UNTIL{ $\left|f(\boldsymbol{\vartheta}_{i+1})-f(\boldsymbol{\vartheta}_i)\right|/f(\boldsymbol{\vartheta}_{i+1})\leq \epsilon_1$, or $i$ reaches the maximum allowable number of iterations.}
\end{algorithmic}
\end{algorithm}

\subsection{Optimizing the FIM Surface Shape $\mathbf{d}$}
\label{sec:surface shape}
In this subsection, the optimization of the surface shape $\mathbf{d}$ is discussed. To address this problem, we propose an efficient PGD optimization algorithm. Since $\mathbf{d}$ appears in the exponential terms of the steering vectors for both the left and right channels of the FIM, directly computing its gradient becomes challenging. Therefore, we first simplify the objective function and define a key intermediate variable, namely, the kernel of the FIM. Subsequently, by vectorizing the kernel, we obtain a more tractable problem. Then, via the chain rule and mathematical derivation, we derive a closed-form expression for the gradient. Finally, we design a closed-form expression for the optimal step size at each iteration and project the deformation in each iteration into the morphing range.

Specifically, by fixing other variables and substituting (18) into (20a) according to (28) and the Appendix, the fundamental form of the optimization problem is given by problem (39). To extract information related to $\mathbf{d}$ more accurately, we rewrite the channels spanning from the $\hat{l}$-th BS to the FIM $\mathbf{G}_{\hat{l}}$ as 

\setcounter{equation}{39}
\begin{equation}
 \mathbf{G}_{\hat{l}}= \mathbf{B}_{\hat{l}}\mathbf{\Gamma}_{\hat{l}}\mathbf{D}_{\hat{l}}^\mathrm{H},
\end{equation}
where $\mathbf{D}_{\hat{l}}\overset{\triangle}{=}[\mathbf{a_{bs}}_{\hat{l},1},\ldots,\mathbf{a_{bs}}_{\hat{l},q},\ldots,\mathbf{a_{bs}}_{\hat{l},Q}]$ $\in \mathbb{C}^{M\times Q}$, and $\mathbf{B}_{\hat{l}}\overset{\triangle}{=}[\mathbf{a_g}_{\hat{l},1},\ldots,\mathbf{a_g}_{\hat{l},q},\ldots,\mathbf{a_g}_{\hat{l},Q}]$ $\in \mathbb{C}^{N\times Q}$. The path gain matrix $\mathbf{\Gamma}_{\hat{l}}$ is defined as $\mathbf{\Gamma}_{\hat{l}}\overset{\triangle}{=}\operatorname{diag}(\beta_{\hat{l},1},\ldots,\beta_{\hat{l},q},\ldots,\beta_{\hat{l},Q})$ $\in \mathbb{C}^{Q\times Q}$. Thus, the cascade channel $\mathbf{H}_{\hat{l},l,k}$ is equivalently translated to 
\begin{align}
 \mathbf{H}_{\hat{l},l,k} &= \operatorname{diag}(\mathbf{h}_{l,k}^r)\mathbf{B}_{\hat{l}}\mathbf{\Gamma}_{\hat{l}}\mathbf{D}_{\hat{l}}^\mathrm{H}\notag\\
&=\mathbf{C}_{\hat{l},l,k}\mathbf{\Gamma}_{\hat{l}}\mathbf{D}_{\hat{l}}^\mathrm{H},
\end{align}
where we define $\mathbf{C}_{\hat{l},l,k}= \operatorname{diag}(\mathbf{h}_{l,k}^r)\mathbf{B}_{\hat{l}}$ $\in \mathbb{C}^{N\times Q}$ as the kernel of the FIM that directly reflects the impact of the surface shape $\mathbf{d}$. 

\noindent\textbf{Remark 2:} The kernel of the FIM $\mathbf{C}_{\hat{l},l,k}$ compactly encapsulates the core physical interaction between the surface shape and the multipath propagation environment. Specifically, its $(n,q)$-th element characterizes how the signal from the $q$-th incident path is passively phase-modulated by the displacement $d_n$ of the $n$-th FIM element, prior to any active phase shift via $\boldsymbol{\vartheta}$.

Furthermore, by merging the matrices unrelated to the kernel of the FIM $\mathbf{C}_{\hat{l},l,k}$ and leaving out constant terms, we can further express the problem (39) as
\begin{align}
&\min_{\mathbf{d}} \mathcal{F}(\mathbf{d})=\sum_{\smash{\hat{l}}=1}^{L} \sum_{l=1}^{L} \sum_{\smash{\hat{k}}=1}^{K}\sum_{k=1}^{K} \boldsymbol{\vartheta}^\mathrm{H}\mathbf{C}_{\hat{l},l,k}\mathbf{f}_{\hat{l},l,\hat{k},k}\mathbf{f}_{\hat{l},l,\hat{k},k}^\mathrm{H}\mathbf{C}_{\hat{l},l,k}^\mathrm{H}\boldsymbol{\vartheta}\notag\\
&\quad\quad\quad\quad+\sum_{\smash{\hat{l}}=1}^{L} \sum_{l=1}^{L} \sum_{\smash{\hat{k}}=1}^{K}\sum_{k=1}^{K} \operatorname{Tr}\left[2\operatorname{Re}\left(\mathbf{T}_{\hat{l},l,\hat{k},k}\mathbf{C}_{\hat{l},l,k}\right)\right]\notag\\
&\quad\quad\quad\quad - \sum_{l=1}^{L} \sum_{k=1}^{K}\operatorname{Tr}\left[2\operatorname{Re}\left(\mathbf{R}_{l,k}\mathbf{C}_{l,l,k}\right)\right]\tag{42a}\\
&\quad\quad\text{s.t.} \quad\quad|d_n| \leq d_{\text{max}}, \quad\forall n = 1, \ldots, N,\tag{42b}
\end{align}
where we have the following definitions:
\setcounter{equation}{42}
\begin{align}
 \mathbf{f}_{\hat{l},l,\hat{k},k}&\overset{\triangle}{=}\sqrt{\omega_{l,k}v_{l,k}\left|u_{l,k}\right|^2}\mathbf{\Gamma}_{\hat{l}}\mathbf{D}_{\hat{l}}^\mathrm{H}\mathbf{w}_{\hat{l},\hat{k}} \in \mathbb{C}^{Q\times 1},\label{eq_43}\\
 \mathbf{T}_{\hat{l},l,\hat{k},k}&\overset{\triangle}{=}\omega_{l,k}v_{l,k}\left|u_{l,k}\right|^2\mathbf{\Gamma}_{\hat{l}}\mathbf{D}_{\hat{l}}^\mathrm{H}\mathbf{w}_{\hat{l},\hat{k}}\mathbf{w}_{\hat{l},\hat{k}}^\mathrm{H}\mathbf{\bar{h}^\mathrm{H}}_{\hat{l},l,k}\boldsymbol{\vartheta}^\mathrm{H} \in \mathbb{C}^{Q\times N},\label{eq_44}\\
 \mathbf{R}_{l,k}&\overset{\triangle}{=}\omega_{l,k}v_{l,k}u_{l,k}^*\mathbf{\Gamma}_l\mathbf{D}_l^\mathrm{H}\mathbf{w}_{l,k}\boldsymbol{\vartheta}^\mathrm{H} \in \mathbb{C}^{Q\times N}.\label{eq_45}
\end{align}

Furthermore, by employing the matrix vectorization method, the first term in (42a) can be further simplified as
\begin{align}
&\quad\boldsymbol{\vartheta}^\mathrm{H}\mathbf{C}_{\hat{l},l,k}\mathbf{f}_{\hat{l},l,\hat{k},k}\mathbf{f}_{\hat{l},l,\hat{k},k}^\mathrm{H}\mathbf{C}_{\hat{l},l,k}^\mathrm{H}\boldsymbol{\vartheta}\notag\\
&=\left|\left|\boldsymbol{\vartheta}^\mathrm{H}\mathbf{C}_{\hat{l},l,k}\mathbf{f}_{\hat{l},l,\hat{k},k}\right|\right|_F^2 =\left|\left|\operatorname{vec}\left(\boldsymbol{\vartheta}^\mathrm{H}\mathbf{C}_{\hat{l},l,k}\mathbf{f}_{\hat{l},l,\hat{k},k}\right)\right|\right|^2\notag\\
&=\left|\left|\left(\mathbf{f}_{\hat{l},l,\hat{k},k}^\mathrm{H}\otimes\boldsymbol{\vartheta}^\mathrm{H} \right)\operatorname{vec}\left(\mathbf{C}_{\hat{l},l,k} \right)\right|\right|^2\notag\\
&=\operatorname{vec}\left(\mathbf{C}_{\hat{l},l,k} \right)^\mathrm{H}\left(\mathbf{f}_{\hat{l},l,\hat{k},k}^\mathrm{H}\otimes\boldsymbol{\vartheta}^\mathrm{H} \right)^\mathrm{H}\left(\mathbf{f}_{\hat{l},l,\hat{k},k}^\mathrm{H}\otimes\boldsymbol{\vartheta}^\mathrm{H} \right)\operatorname{vec}\left(\mathbf{C}_{\hat{l},l,k} \right)\notag\\
&=\mathbf{y}_{\hat{l},l,k}^\mathrm{H}\mathbf{A}_{\hat{l},l,\hat{k},k}\mathbf{y}_{\hat{l},l,k},
\end{align}
where $\mathbf{y}_{\hat{l},l,k}=\operatorname{vec}\left(\mathbf{C}_{\hat{l},l,k} \right) \in \mathbb{C}^{NQ}$, and $\mathbf{A}_{\hat{l},l,\hat{k},k}$ is given by
\begin{equation}
\mathbf{A}_{\hat{l},l,\hat{k},k}=\left(\mathbf{f}_{\hat{l},l,\hat{k},k}^\mathrm{H}\otimes\boldsymbol{\vartheta}^\mathrm{H} \right)^\mathrm{H}\left(\mathbf{f}_{\hat{l},l,\hat{k},k}^\mathrm{H}\otimes\boldsymbol{\vartheta}^\mathrm{H} \right)\in \mathbb{C}^{NQ\times NQ}.
\end{equation}

In addition, the other terms in (42a) can be further formulated as
\begin{equation}
\operatorname{Tr}[2\operatorname{Re}(\mathbf{T}_{\hat{l},l,\hat{k},k}\mathbf{C}_{\hat{l},l,k})]=2\operatorname{Re}\left(\mathbf{t}_{\hat{l},l,\hat{k},k}^\mathrm{H} \mathbf{y}_{\hat{l},l,k}\right), 
\end{equation}
\begin{equation}
\operatorname{Tr}[2\operatorname{Re}(\mathbf{R}_{l,k}\mathbf{C}_{l,l,k})]=2\operatorname{Re}\left(\mathbf{r}_{l,k}^\mathrm{H} \mathbf{y}_{l,l,k}\right), 
\end{equation}
where we define $\mathbf{t}_{\hat{l},l,\hat{k},k}=\operatorname{vec}\left(\mathbf{T}_{\hat{l},l,\hat{k},k}^\mathrm{H}\right)\in \mathbb{C}^{NQ}$, and $\mathbf{r}_{l,k}=\operatorname{vec}\left(\mathbf{R}_{l,k}^\mathrm{H}\right)\in \mathbb{C}^{NQ}$. By doing so, we can further express the FIM surface shape optimization problem (41) as 
\begin{align}
&\min_{\mathbf{d}} \mathcal{F}(\mathbf{d})=\sum_{\smash{\hat{l}}=1}^{L} \sum_{l=1}^{L} \sum_{\smash{\hat{k}}=1}^{K}\sum_{k=1}^{K} \mathbf{y}_{\hat{l},l,k}^\mathrm{H}\mathbf{A}_{\hat{l},l,\hat{k},k}\mathbf{y}_{\hat{l},l,k}\notag\\
&\quad\quad\quad\quad+\sum_{\smash{\hat{l}}=1}^{L} \sum_{l=1}^{L} \sum_{\smash{\hat{k}}=1}^{K}\sum_{k=1}^{K} 2\operatorname{Re}\left(\mathbf{t}_{\hat{l},l,\hat{k},k}^\mathrm{H} \mathbf{y}_{\hat{l},l,k}\right)\notag\\
&\quad\quad\quad\quad - \sum_{l=1}^{L} \sum_{k=1}^{K}2\operatorname{Re}\left(\mathbf{r}_{l,k}^\mathrm{H} \mathbf{y}_{l,l,k}\right)\tag{50a}\\
&\quad\quad\text{s.t.} \quad\quad|d_n| \leq d_{\text{max}}, \quad\forall n = 1, \ldots, N.\tag{50b}
\end{align}

Due to the complex exponential term of the kernel vector $\mathbf{y}_{\hat{l},l,k}$ with respect to surface shape $\mathbf{d}$, the expression of problem (50) is not a convex problem. However, it is clear that constraint (50b) is a convex linear constraint, and the objective function (50a) has a concise form, making its gradient computable. Therefore, to find the optimal FIM surface shape, we utilize a projected gradient descent (PGD) algorithm including the following three steps.

\begin{figure*}[t]
\vspace*{-0.5cm} 
\begin{equation}
\frac{\partial \mathcal{F}(\mathbf{d})}{\partial d_n} = \sum_{\smash{\hat{l}}=1}^{L} \sum_{l=1}^{L} \sum_{\smash{\hat{k}}=1}^{K}\sum_{k=1}^{K} 2 \operatorname{Re} \left\{\mathbf{y}_{\hat{l},l,k}^H \mathbf{A}_{\hat{l},l,\hat{k},k} \left(\frac{\partial \mathbf{y}_{\hat{l},l,k}}{\partial d_n}\right)+\mathbf{t}_{\hat{l},l,\hat{k},k}^\mathrm{H} \left(\frac{\partial \mathbf{y}_{\hat{l},l,k}}{\partial d_n}\right)\right\}- \sum_{l=1}^{L} \sum_{k=1}^{K}2\operatorname{Re}\left\{\mathbf{r}_{l,k}^\mathrm{H}\left(\frac{\partial \mathbf{y}_{l,l,k}}{\partial d_n}\right)\right\}.\tag{51}
\end{equation}
\begin{equation}
{\mathcal{H}_1(\rho)}_{\hat{l},l,\hat{k},k}=\sum_{i=1}^
{NQ}a^{i,i}_{\hat{l},l,\hat{k},k}+2\operatorname{Re}\left[\sum_{i=1}^{NQ}\sum_{\smash{\hat{i}}>i}^{NQ}a^{i,\hat{i}}_{\hat{l},l,\hat{k},k}e^{j\left(\xi_{\hat{l},l,k}^{i(t)}-\xi_{\hat{l},l,k}^{\hat{i}(t)} \right)}e^{j\rho\left(\chi_{\hat{l},l,k}^{\hat{i}(t)}-\chi_{\hat{l},l,k}^{i(t)} \right)}+\sum_{i=1}^{NQ}t_{\hat{l},l,\hat{k},k}^{i}e^{j\xi_{\hat{l},l,k}^{i(t)}}e^{j\rho\chi_{\hat{l},l,k}^{i(t)}}\right].\tag{62}
\end{equation}
\hrulefill
\end{figure*}

\subsubsection{Compute Euclidean Gradient}The derivative of (50a) with respect to $d_n$ can be calculated as (51). We first rewrite the kernel of the FIM $\mathbf{C}_{\hat{l},l,k}$ by performing block row partitioning on $\mathbf{B}_{\hat{l}}$ and block column partitioning on $\operatorname{diag}(\mathbf{h}_{l,k}^r)$. By doing so, we have $\mathbf{B}_{\hat{l}}=[\mathbf{b}_{\hat{l},1},\ldots,\mathbf{b}_{\hat{l},i},\ldots,\mathbf{b}_{\hat{l},N}]^\mathrm{H}$, and $\mathbf{b}_{\hat{l},i}=[b_{\hat{l},i,1},\ldots,b_{\hat{l},i,q},\ldots,b_{\hat{l},i,Q}]^\mathrm{H} \in \mathbb{C}^{Q\times 1}$, where
\setcounter{equation}{51}
\begin{equation}
 b_{\hat{l},i,q}= \delta_{\hat{l},i,q} e^{j\kappa d_n\cos{\theta^I_{\hat{l},q}}\cos{\phi^I_{\hat{l},q}}},
\end{equation}
and $\delta_{\hat{l},i,q}$ is the $i$-th element of the steering vector of the unmorphed FIM $\mathbf{a}_{\text{upa}}(\theta_{\hat{l},q}^I,\phi_{\hat{l},q}^I)$. Meanwhile, we have
\begin{equation}
 \operatorname{diag}(\mathbf{h}_{l,k}^r)=\operatorname{diag}\left( \sum_{p=1}^{P}\mathbf{a}_{\mathbf{h}_{l,k,p}}\right) = \sum_{p=1}^{P}\operatorname{diag}\left( \mathbf{a}_{\mathbf{h}_{l,k,p}}\right),
\end{equation}
and $\operatorname{diag}\left( \mathbf{a}_{\mathbf{h}_{l,k,p}}\right)=[\mathbf{e}_{l,k,p,1},\ldots,\mathbf{e}_{l,k,p,i},\ldots,\mathbf{e}_{l,k,p,N}]$. Thus, the kernel of the FIM $\mathbf{C}_{\hat{l},l,k}$ can be derived as 
\begin{equation}
\mathbf{C}_{\hat{l},l,k}=\sum_{p=1}^{P}\sum_{i=1}^{N}\mathbf{e}_{l,k,p,i}\mathbf{b}_{\hat{l},i}^\mathrm{H}.
\end{equation}

It can be observed from (54) that the kernel of the FIM is the sum of the interactions between the deformation distance $d_n$ of each element of the FIM and the multipath effects, indicating that the morphing surface shape of the FIM can effectively reconfigure the radio environments. The derivative of $\mathbf{y}_{\hat{l},l,k}$ with respect to $d_n$ yields
\begin{align}
 &\frac{\partial \mathbf{y}_{\hat{l},l,k}}{\partial d_n}=\operatorname{vec}\left[\sum_{p=1}^{P}\sum_{i=1}^{N}\left( \frac{\partial \mathbf{e}_{l,k,p,i}}{\partial d_n}\mathbf{b}_{\hat{l},i}^\mathrm{H} + \mathbf{e}_{l,k,p,i}\frac{\partial \mathbf{b}_{\hat{l},i}^\mathrm{H}}{\partial d_n} \right)\right]\notag\\
 &=\operatorname{vec}\left[\sum_{p=1}^{P}\sum_{i=1}^{N}\left(\zeta_{l,k,p}\mathbf{e}_{l,k,p,i}\mathbf{b}_{\hat{l},i}^\mathrm{H} +\mathbf{e}_{l,k,p,i}\mathbf{b}_{\hat{l},i}^\mathrm{H}\mathbf{Q}_{\hat{l}}\right)\right],
\end{align}
where $\zeta_{l,k,p}$ and $\mathbf{Q}_{\hat{l}} \in \mathbb{C}^{Q\times Q}$ are given by
\begin{equation}
 \zeta_{l,k,p}= j\kappa\cos{\theta^O_{l,k,p}}\cos{\phi^O_{l,k,p}},
\end{equation}
\begin{equation}
 \mathbf{Q}_{\hat{l}}=\operatorname{diag}\left(j\kappa \cos{\theta^I_{\hat{l},q}}\cos{\phi^I_{\hat{l},q}},\ldots,j\kappa \cos{\theta^I_{\hat{l},Q}}\cos{\phi^I_{\hat{l},Q}}\right).
\end{equation}

Therefore, an accurate expression for $\partial \mathcal{F}(\mathbf{d})/{\partial d_n}$ can be obtained.
\begin{figure*}[t]
\vspace{-0.5cm}
\begin{equation}
 c_1=\sum_{\smash{\hat{l}}=1}^{L} \sum_{l=1}^{L} \sum_{\smash{\hat{k}}=1}^{K}\sum_{k=1}^{K}\operatorname{Re}\left[\sum_{i=1}^{NQ}\sum_{\smash{\hat{i}}>i}^{NQ}a_{\hat{l},l,\hat{k},k}^{i,\hat{i}}e^{j\left(\xi_{\hat{l},l,k}^{i(t)}-\xi_{\hat{l},l,k}^{\hat{i}(t)} \right)}\left(\chi_{\hat{l},l,k}^{\hat{i}(t)}-\chi_{\hat{l},l,k}^{i(t)}\right)+\sum_{i=1}^{NQ}jt_{\hat{l},l,\hat{k},k}^{i}e^{j\xi_{\hat{l},l,k}^{i(t)}}\chi_{\hat{l},l,k}^{i(t)}\right],\tag{67}
\end{equation}
\begin{equation}
 c_2=\sum_{\smash{\hat{l}}=1}^{L} \sum_{l=1}^{L} \sum_{\smash{\hat{k}}=1}^{K}\sum_{k=1}^{K}\operatorname{Re}\left[\sum_{i=1}^{NQ}\sum_{\smash{\hat{i}}>i}^{NQ}a_{\hat{l},l,\hat{k},k}^{i,\hat{i}}e^{j\left(\xi_{\hat{l},l,k}^{i(t)}-\xi_{\hat{l},l,k}^{\hat{i}(t)} \right)}\left(\chi_{\hat{l},l,k}^{\hat{i}(t)}-\chi_{\hat{l},l,k}^{i(t)}\right)^2+\sum_{i=1}^{NQ}t_{\hat{l},l,\hat{k},k}^{i}e^{j\xi_{\hat{l},l,k}^{i(t)}}\left(\chi_{\hat{l},l,k}^{i(t)}\right)^2\right].\tag{68}
\end{equation}
\hrulefill
\end{figure*}

\subsubsection{Find the Suitable Step Size}
After obtaining the descent direction of the objective function, a suitable step size $\rho > 0$ needs to be determined. To this end, denote by $\mathbf{g}^{(t)}=\nabla_\mathbf{d} \mathcal{F}(\mathbf{d})^{(t)}$ the adopted descent direction at iteration $t$, and by $\mathbf{y}_{\hat{l},l,k}^\mathrm{(t)}$ the vectorized kernel at the $t$-th iteration $t$. Then, the next iteration point can be obtained by
\begin{equation}
 \mathbf{d}^{(t+1)}=\mathbf{d}^{(t)}+\rho\mathbf{g}^{(t)},
\end{equation}
\begin{equation}
 \mathbf{y}_{\hat{l},l,k}^{(t+1)}=\mathbf{y}_{\hat{l},l,k}^{(t)}\odot e^{j\rho\boldsymbol{\chi}_{\hat{l},l,k}^{(t)}},
\end{equation}
where we define $\rho\boldsymbol{\chi}_{\hat{l},l,k}^{(t)}$ as the phase difference between $\mathbf{y}_{\hat{l},l,k}^{(t)}$ and $\mathbf{y}_{\hat{l},l,k}^{(t+1)}$, and $\boldsymbol{\xi}_{\hat{l},l,k}^{(t)}$ as the phase of $\mathbf{y}_{\hat{l},l,k}^{(t)}$.

Furthermore, an appropriate step size can be obtained by solving the following minimization problem:
\begin{align}
&\min_{\rho>0} \mathcal{H}(\rho)=\sum_{\smash{\hat{l}}=1}^{L} \sum_{l=1}^{L} \sum_{\smash{\hat{k}}=1}^{K}\sum_{k=1}^{K} \left(\mathbf{y}_{\hat{l},l,k}^{(t+1)}\right)^\mathrm{H}\mathbf{A}_{\hat{l},l,\hat{k},k}\mathbf{y}_{\hat{l},l,k}^{(t+1)}\notag\\
&\quad\quad\quad\quad+\sum_{\smash{\hat{l}}=1}^{L} \sum_{l=1}^{L} \sum_{\smash{\hat{k}}=1}^{K}\sum_{k=1}^{K} 2\operatorname{Re}\left(\mathbf{t}_{\hat{l},l,\hat{k},k}^\mathrm{H} \mathbf{y}_{\hat{l},l,k}^{(t+1)}\right)\notag\\
&\quad\quad\quad\quad - \sum_{l=1}^{L} \sum_{k=1}^{K}2\operatorname{Re}\left(\mathbf{r}_{l,k}^\mathrm{H} \mathbf{y}_{l,l,k}^{(t+1)}\right)\tag{60a}\\
&\quad\quad\text{s.t.} \quad\quad|d_n| \leq d_{\text{max}}, \quad\forall n = 1, \ldots, N.\tag{60b}
\end{align}

Furthermore, we expand $\mathcal{H}(\rho)$ by plugging (58) and (59) into (60a) and obtain
\setcounter{equation}{60}
\begin{equation}
\mathcal{H}(\rho)=\sum_{\smash{\hat{l}}=1}^{L} \sum_{l=1}^{L} \sum_{\smash{\hat{k}}=1}^{K}\sum_{k=1}^{K}{\mathcal{H}_1(\rho)}_{\hat{l},l,\hat{k},k}-\sum_{l=1}^{L} \sum_{k=1}^{K}{\mathcal{H}_2(\rho)}_{l,k}.
\end{equation}
\begin{algorithm}[t]
\label{a2}
\renewcommand{\algorithmicrequire}{\textbf{Input:}}
\renewcommand{\algorithmicensure}{\textbf{Output:}}
\renewcommand{\algorithmicrepeat}{\textbf{Repeat}}
\renewcommand{\algorithmicuntil}{\textbf{Until}}
\renewcommand{\algorithmicif}{\textbf{if}}
\caption{Projected Gradient Descent Algorithm (PGD)}
\begin{algorithmic}[1]
\STATE Initialize the iteration number $t$ = 0 and the accuracy $\epsilon_2$. 
\STATE Construct a feasible solution $\mathbf{d}^{(0)}$.
\REPEAT
 \STATE Calculate the vectorized FIM kernel $\mathbf{y}_{\hat{l},l,k}^{(t)}$ according to (54).
 \STATE Calculate the objective function $\mathcal{F}(\mathbf{d})^{(t)}$ in (50a).
 \STATE Calculate the gradient $\mathbf{g}^{(t)}$ according to (51) and (55).
 \STATE Calculate $\upsilon_1$ in (65) and $\upsilon_2$ in (66).
 \IF{ $\upsilon_1 \geq 0$, $\upsilon_2 > 0$ }
 \STATE Set the step size as $\rho = \upsilon_1/2\upsilon_2$.
 \ELSE
 \STATE Choose the step size $\rho$ by applying the linear search method.
 \ENDIF
 \STATE Update the FIM surface shape by $\mathbf{d}^{(t+1)}=\mathbf{d}^{(t)}+\rho\mathbf{g}^{(t)}$.
 \STATE Project the surface shape to the morphing range according to $d_n^{(t+1)} = \max(\min(d_n^{(t+1)},d_{\text{max}}),-d_{\text{max}})$.
 \STATE Update the iteration counter by $t+1\to t$.
\UNTIL{ $|\mathcal{F}(\mathbf{d})^{(t+1)}-\mathcal{F}(\mathbf{d})^{(t)}|/\mathcal{F}(\mathbf{d})^{(t+1)}\leq \epsilon_2$, or $t$ exceeds the maximum tolerable number of iterations.}
\end{algorithmic}
\end{algorithm}
The detailed expression of ${\mathcal{H}_1(\rho)}_{\hat{l},l,\hat{k},k}$ is shown in (62), and 
\setcounter{equation}{62}
\begin{equation}{\mathcal{H}_2(\rho)}_{l,k}=2\operatorname{Re}\left[\sum_{i=1}^{NQ}r_{l,k}^ie^{j\xi_{l,l,k}^{i(t)}}e^{j\rho\chi_{l,l,k}^{i(t)}}\right],
\end{equation} 
where $a^{i,\hat{i}}_{\hat{l},l,\hat{k},k}$ is the $(i,\hat{i})$-th element of matrix $\mathbf{A}_{\hat{l},l,\hat{k},k}$. $\xi_{\hat{l},l,k}^{i(t)}$, $\chi_{\hat{l},l,k}^{i(t)}$ and $r_{l,k}^i$ denote the $i$-th element of vector $\boldsymbol{\xi}_{\hat{l},l,k}^{(t)}$, $\boldsymbol{\chi}_{\hat{l},l,k}^{(t)}$ and $\mathbf{r}_{l,k}^\mathrm{H}$, respectively. In addition, $t_{\hat{l},l,\hat{k},k}^{i}$ is the $i$-th element of vector $\mathbf{t}_{\hat{l},l,\hat{k},k}^\mathrm{H}$.

In fact, a linear search method can be employed to determine $\rho > 0$ that minimizes $\mathcal{H}(\rho)$. In order to reduce complexity, we can approximate the objective function $\mathcal{H}(\rho)$ using its second-order Maclaurin expansion, thereby obtaining
\begin{equation}
 \hat{\mathcal{H}}(\rho)= \upsilon_0 + 2\upsilon_1\rho -2\upsilon_2\rho^2.
\end{equation}
Thus, the approximation $\hat{\mathcal{H}}(\rho)$ admits an optimal closed-form step size $\rho^*=\upsilon_1/2\upsilon_2$ if $\upsilon_1 \geq 0$ and $\upsilon_2 >0$. The expression of $\upsilon_1$ and $\upsilon_2$ are derived as
\begin{equation}
 \upsilon_1=c_1-\sum_{l=1}^{L}\sum_{k=1}^{K}\operatorname{Re}\left[\sum_{i=1}^{NQ}jr_{l,k}^ie^{j\xi_{l,l,k}^{i(t)}}\chi_{l,l,k}^{i(t)}\right],
\end{equation}
\begin{equation}
 \upsilon_2=c_2-\sum_{l=1}^{L}\sum_{k=1}^{K}\operatorname{Re}\left[\sum_{i=1}^{NQ}r_{l,k}^ie^{j\xi_{l,l,k}^{i(t)}}\left(\chi_{l,l,k}^{i(t)}\right)^2\right],
\end{equation}
where $c_1$ and $c_2$ are given in (67) and (68), respectively. It should be mentioned that if the above conditions do not hold, traditional linear search methods can be used.

\subsubsection{Projection}
At each iteration, equipped with a closed-form expression for calculating the gradient of the objective function and a suitable step size, a projection process is applied to the elements of the surface shape $\mathbf{d}$ in (58) to satisfy the morphing range constraint, resulting in
\setcounter{equation}{68}
\begin{equation}
 d_n = \max\left(\min\left(d_n,d_{\text{max}}\right),-d_{\text{max}}\right).
\end{equation}

The details of the PGD algorithm for optimizing the FIM surface shape are presented in Algorithm 2.

\subsection{Overall Block Coordinate Descent Algorithm}
\label{subsec:overall}
Based on the above discussion, we present the detailed procedure of the proposed WMMSE and BCD-based WSR optimization algorithm in Algorithm~3. In this subsection, we will discuss the algorithm's convergence and complexity analysis, as well as examine its scalability and real-time scenario performance.
\subsubsection{Convergence Analysis} The convergence of the proposed algorithm is ensured by two key factors. Firstly, in each subproblem, at least a locally optimal solution is guaranteed, ensuring that the overall objective function is non-decreasing throughout the iterative process, assuming (16a) is the overall objective. Secondly, the three primary optimization variables are all continuous and subject to well-defined constraints on transmission power, morphing range, and unit-modulus reflection coefficients. These constraints inherently establish an upper bound for the objective function, thereby guaranteeing the convergence.
\begin{algorithm}[t]
\label{a2}
\renewcommand{\algorithmicrequire}{\textbf{Input:}}
\renewcommand{\algorithmicensure}{\textbf{Output:}}
\renewcommand{\algorithmicrepeat}{\textbf{Repeat}}
\renewcommand{\algorithmicuntil}{\textbf{Until}}
\renewcommand{\algorithmicif}{\textbf{if}}
\caption{Block Coordinate Descent Algorithm (BCD) }
\begin{algorithmic}[1]
\STATE Initialize the iteration number $t$ = 0, feasible solutions $\mathbf{d}^{(0)}$, $\boldsymbol{\Phi}^{(0)}$ and $\mathbf{W}^{(0)}$.
\REPEAT
 \STATE Calculate the decoding vectors $\mathbf{u}^{(t)}$ by (21).
 \STATE Calculate the auxiliary matrices $\mathbf{v}^{(t)}$ by (22).
 \STATE Update the beamforming vectors $\mathbf{W}^{(t+1)}$ by (25).
 \STATE Update the phase shift matrix $\boldsymbol{\Phi}^{(t+1)}$ using the RCG algorithm developed in Section \ref{sec:phase shift};
 \STATE Update the FIM surface shape vector $\mathbf{d}^{(t+1)}$ by using the PGD algorithm in Section 
\ref{sec:surface shape}.
 \STATE Update the iteration counter by $t+1\to t$.
\UNTIL{Objective function in (16a) converges, or $t$ exceeds the maximum tolerable number of iterations.}
\end{algorithmic}
\end{algorithm}
\subsubsection{Complexity Analysis} To analyze the overall complexity of Algorithm 3, we need to discuss the complexity of each block in the BCD algorithm with $L_{\text{BCD}}$ iterations. The complexity of optimizing the decoding vector $\mathbf{u}$ and the auxiliary vector $\mathbf{v}$ are given by $\mathcal{O}(LKM^3)$ and $\mathcal{O}(LK)$, respectively. In Step 5, the computational cost of beamforming optimization is mainly dominated by the matrix inversion operation, which results in a complexity of $\mathcal{O}(LKM^3)$. In Step 6, we get the optimal phase shift matrix using the RCG algorithm, where the worst-case total computational complexity is $\mathcal{O}(L_{\text{RCG}} N^{1.5})$. In Step 7, the complexity of computing the optimal FIM surface shape vector is mainly concentrated in gradient calculation and step size determination. The gradient computation primarily involves the multiplication of the vectorized multi-path channel matrix, yielding a complexity of $\mathcal{O}(LKPQN)$. Additionally, the computing complexity for the step size is $\mathcal{O}(N)$. Assuming that the PGD algorithm performs $L_{\text{PGD}}$ iterations, the overall complexity of the PGD algorithm is $\mathcal{O}(L_{\text{PGD}} LKPQN)$. It can be observed that the computational complexity of the PGD algorithm increases linearly with the number of FIM elements $N$. Therefore, the complexity of the overall BCD algorithm is $\mathcal{O} \left( L_{\text{BCD}} \left( LKM^3 + LK + L_{\text{RCG}} N^{1.5} + L_{\text{PGD}} LKPQN \right) \right)$.

\subsubsection{Algorithm Scalability}The core of the algorithm's scalability lies in the fact that the dimension of the optimization variables $\mathbf{d}$ and $\boldsymbol{\vartheta}$ is fixed at $N$, independent of the network scale. Moreover, due to the summation form in the objective function and gradient computation, an increase in network scale only leads to an increase in the number of summation terms. This computation inherently possesses ``data-parallel" characteristics and the contribution of each user pair can be solved at each BS independently, providing direct feasibility for parallel computing. Therefore, the algorithm can handle ultra-dense deployment through parallelization.

\subsubsection{Real-time Scenario Performance}The proposed algorithm demonstrates inherent potential for real-time operation due to three key factors. First, its computational complexity is dominated by the phase shift optimization (scaling with $N^{1.5}$), which is a process already proven feasible within channel coherence time in mature RIS literature\cite{ref21,ref54}. Second, the response time of FIM surface morphing reaches 10 ms according to Table~I and aligns with typical channel coherence blocks, providing a sufficient time window for joint optimization. Third, for practical scenarios with fast time-varying channels, we can also adopt a dual time-scale approach, trading a marginal degree of precision for higher response speed. Specifically, phase shift configuration and surface shape configuration can operate on two different time scales: (I) Phase shifts can be optimally configured within each channel coherence interval. (II) The FIM surface shape can be reconfigured every few time slots based on statistical CSI aggregated over multiple coherence intervals.

\subsubsection{Practical Configuration of FIM}The proposed BCD algorithm is assumed to be implemented in a centralized manner. During the optimization process, the physical FIM remains unchanged, while the algorithm iteratively computes the optimal configuration. Upon convergence, the optimized parameters are conveyed to the BS and the FIM controller, enabling a simultaneous, one-step configuration. However, in practical scenarios, given that the deformation response of FIMs operates on the order of milliseconds (with the fastest reaching 1 ms) [38]-[40], while the phase-shifting response time lies within microseconds (nano-second level is achievable for varactor or Positive-Intrinsic-Negative (PIN) diode-based schemes)\cite{ref52}, a temporal mismatch may occur during the joint configuration of FIM phase shifts and shape morphing. This mismatch can lead to dynamic misalignment issues, resulting in brief periods of uncontrollable performance. In-depth investigation of this challenge will constitute a significant research direction in the future development of FIM technology.

\section{Simulation Results}
\label{sec:Simulation Result}
In this section, numerical results are provided to validate the effectiveness of the proposed algorithm and to draw key insights into the performance of FIM-aided systems. The following subsections will present the setup of the simulation experiments and an analysis of the performance.
\subsection{Simulation Setup}
As shown in Fig. 3, an FIM-aided downlink multicell MU-MISO network is considered. The FIM is composed of $N=N_yN_z$ EM units. For clarity of exposition, $N_z$ is fixed at 2, while $N_y$ is increased linearly. Additionally, we consider a two-cell ($L=2$) scenario, where each micro cell is modeled as a hexagonal region with a circumscribed radius of 90 m,  centered at coordinates (0 m, 200 m, 30 m) and (0 m, 400 m, 30 m), respectively. Each BS, equipped with $M=4$ TAs, is deployed at the center of its respective elliptical cell. An FIM is positioned at the intersection of the two cells, with coordinates set at (0 m, 300 m, 30 m). We set $K=2$ edge UEs in each cell, randomly distributed on both sides of the FIM. Additionally, we consider randomly placed obstacles acting as scatterers between the BSs and UEs, creating a complex propagation environment. Moreover, we introduce the large-scale path loss model as
\begin{equation}
 Loss = Loss_0-10\ell\log_{10}\left(\frac{d}{d_0}\right),
\end{equation}
where $Loss_0= -30$ dB is the path loss at the reference propagation distance $d_0=1$ m and $\ell$ is the path loss exponent. We assume that the reflection path loss $\alpha_{l,k,p}$ and $\beta_{\hat{l},q}$ are determined by $\ell_{\text{BF}} = \ell_{\text{FU}} = 2.2$ and the direct path loss exponent is given by $\ell_{\text{BU}}=3.67$. We assume that the azimuth and elevation angles on both sides of the FIM, denoted as $\theta_{\hat{l},q}^I, \phi_{\hat{l},q}^I, \theta_{l,k,p}^O, \phi_{l,k,p}^O$, along with the transmit angle at the BS, $\gamma_{\hat{l},q}$, follow a uniform distribution over the range of $[-\pi/2, \pi/2)$. The system is configured to operate at a wavelength of $\lambda = 0.01$ m. The noise power is $\sigma_{l,k}^2 =\sigma^2= -80$ dBm. Moreover, we assume the same transmit power $P_{t,l}=P_t$ at different BSs, without loss of generality. Weighting factors $\omega_{l,k}$ are assigned inversely proportional to the average path loss between each user and its serving BS, subsequently normalized by $\sum_{l=1}^L\sum_{k=1}^K \omega_{l,k} = LK$. Unless otherwise stated, the the number of NLoS paths is defaulted as $P=Q=2$. Additionally, we assume that FIM can continuously morph its shape within the morphing range $d_{\text{max}}$\footnote{In practical implementations, the deformation error of the FIM is typically much smaller than the operating wavelength and can therefore be safely ignored. In addition, the deformation reconfiguration time of the FIM is on the millisecond scale, which is comparable to the channel coherence time, enabling timely adaptation to the wireless channel variations.}, which is defaulted to $\lambda$. The stopping threshold of the RCG and PGD algorithms are set as $\epsilon_1 =\epsilon_2= 10^{-4}$. The maximum allowable number of iterations is set to $500$. The initial values are set as $\mathbf{d}^{(0)} = \mathbf{0}$, the phase of $\boldsymbol{\vartheta}_n^{(0)}$, namely $\varphi_n$ is randomly distributed within $[0, 2\pi]$, and $\mathbf{w}^{(0)}$ is set based on normalized zero-forcing (ZF) precoding.

In addition, to model the imperfect CSI scenario and to verify the robustness of the proposed algorithm, we introduce the normalized MSE $\varsigma$ of the parameters to be estimated. Specifically, under the channel modeling considered in this paper, the main parameters that need to be estimated include the direct channel $\mathbf{\bar{h}}_{\hat{l},l,k}$ as well as the azimuth and elevation angles $\phi_{l,k,p}^O$, $\theta_{l,k,p}^O$, $\phi_{\hat{l},q}^I$, $\theta_{\hat{l},q}^I$. Denote $x$ and $\hat{x}$ as the actual value and estimated value of the parameters, respectively. Assuming that the estimation error follows a zero-mean Gaussian distribution, all parameters are modeled to share the same normalized MSE $\varsigma$ as 
\begin{equation}
\varsigma
= \frac{\mathbb{E}\!\left[ \lvert x - \hat{x} \rvert^{2} \right]}
{\mathbb{E}\!\left[ \lvert \hat{x} \rvert^{2} \right]},
\end{equation}
which directly reflects the accuracy of channel estimation.
\begin{figure}[t] 
\centering %
\includegraphics[width=0.42\textwidth]{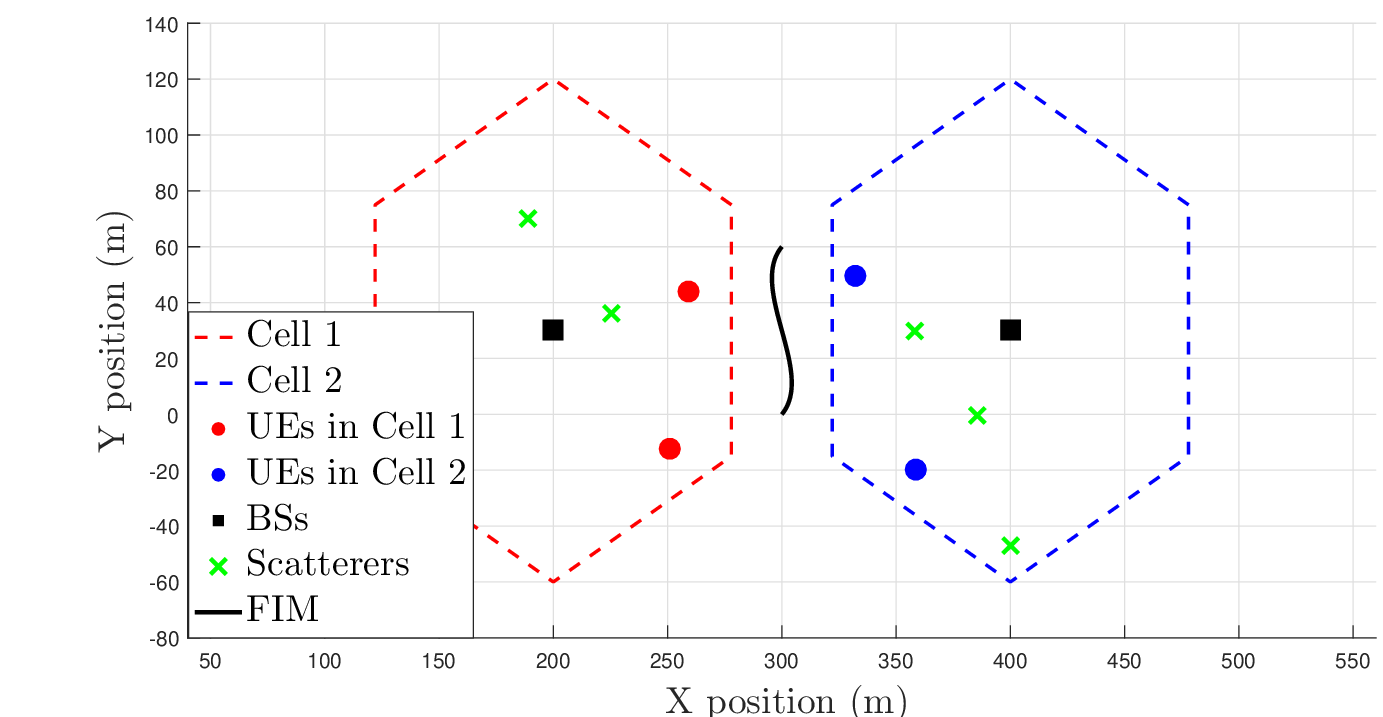}
\caption{The simulated FIM-aided MU-MISO communication scenario, where two cells are considered.} 
\label{fig_1}
\vspace{-0.5cm}
\end{figure}
\begin{figure}[t] 
\centering %
\includegraphics[width=0.42\textwidth]{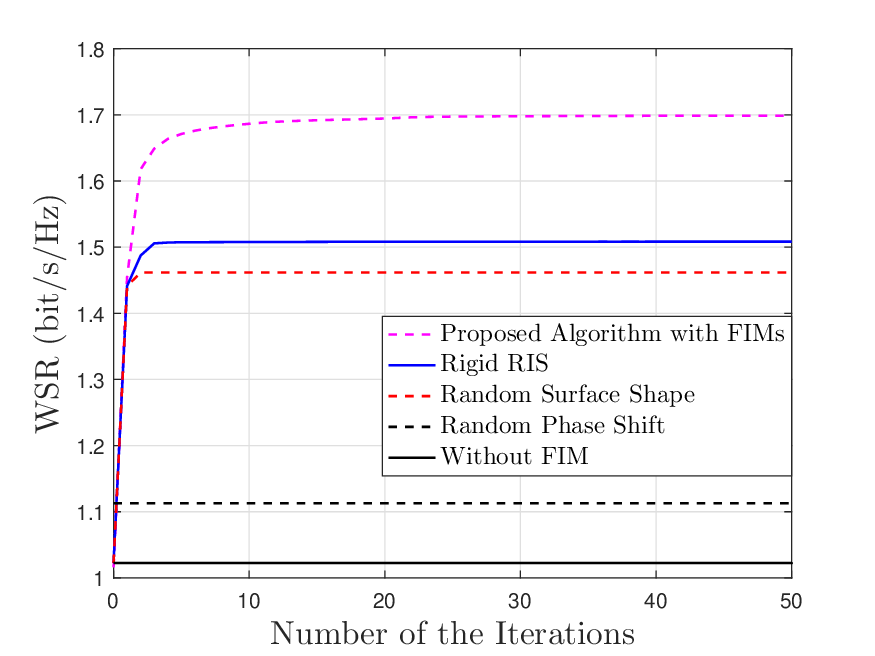}
\caption{Convergence behavior of WSR when $N=30$ and $P_t=30$ dBm.}
\label{fig_1}
\vspace{-0.5cm}
\end{figure}

\begin{figure}[!t]
\centering
\subfloat[]{\includegraphics[width=0.234\textwidth]{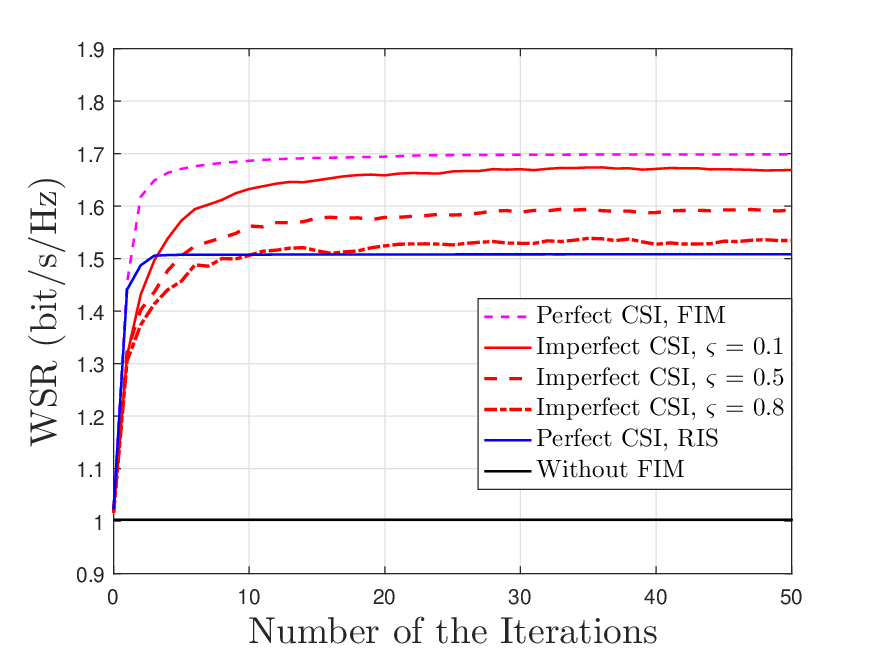}}
\hfill
\subfloat[]{\includegraphics[width=0.234\textwidth]{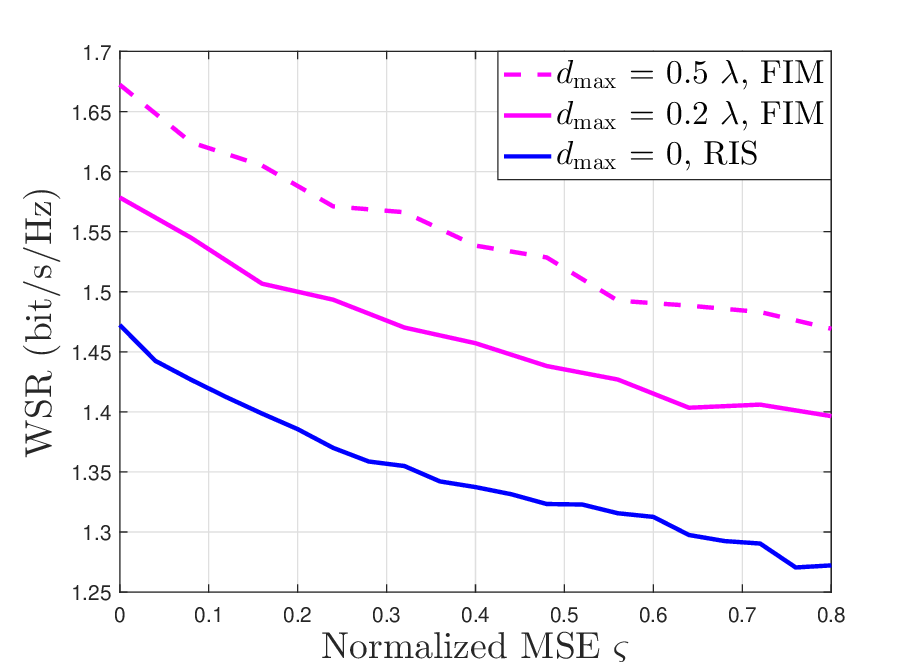}}
		\caption{(a) Convergence behavior under different values of normalized MSE $\varsigma$; (b) WSR versus normalized MSE $\varsigma$}
\label{fig_3}
\vspace{-0.5cm}
\end{figure}

\begin{figure}[!t]
\centering
\subfloat[]{\includegraphics[width=0.234\textwidth]{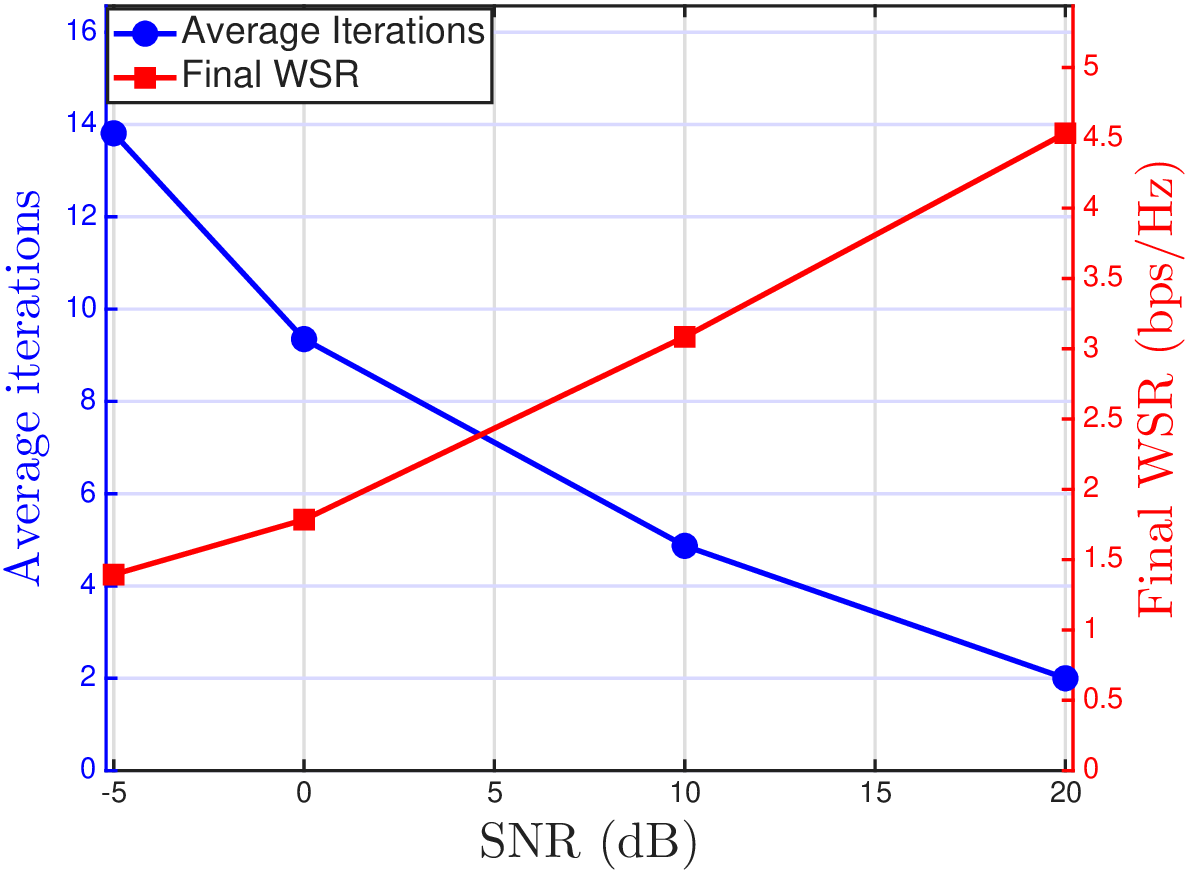}}
\hfill
\subfloat[]{\includegraphics[width=0.234\textwidth]{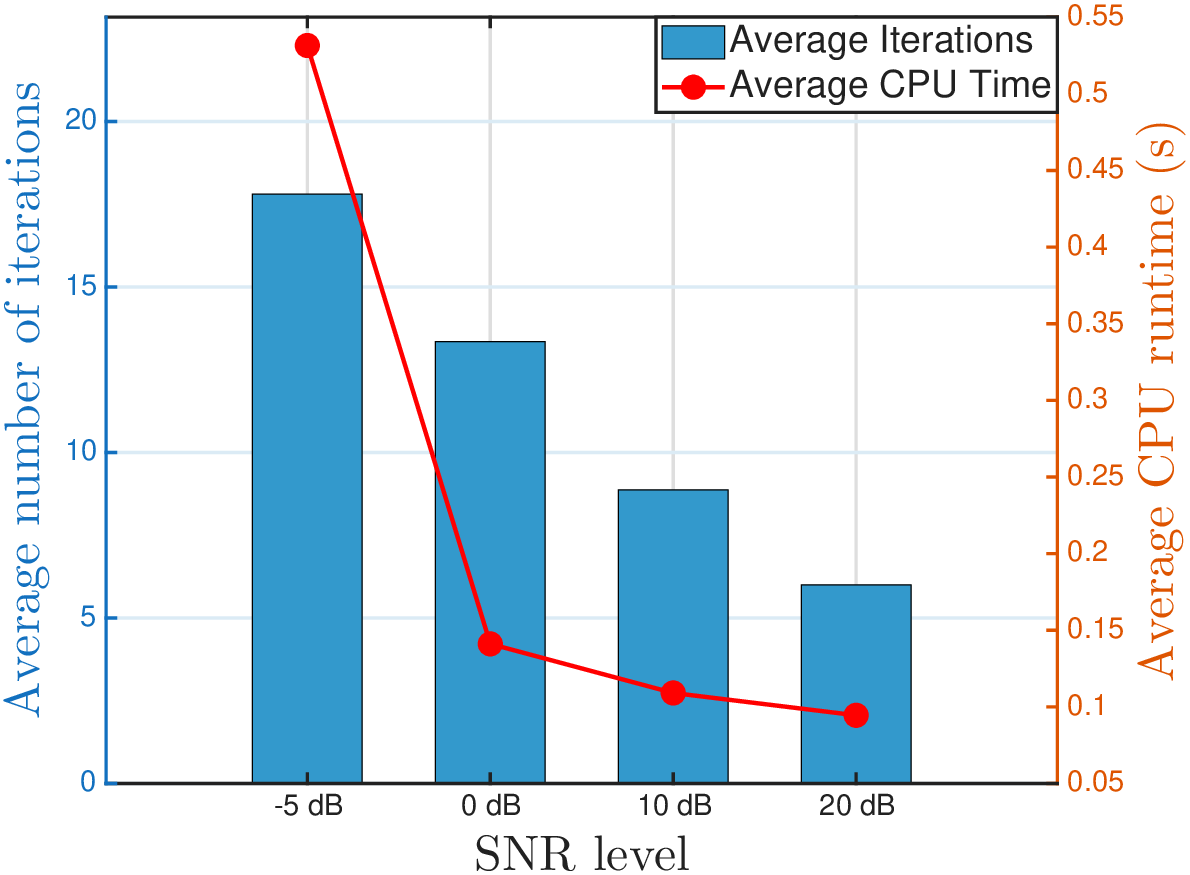}}
\caption{(a) Relationship between WSR, average iterations and SNR; (b) Relationship between SNR, average iterations and runtime.}
\label{fig_3}
\vspace{-0.5cm}
\end{figure}

To investigate the efficacy of the proposed algorithms, the following 4 baselines are considered:
\begin{itemize}
 \item \textbf{Rigid RIS}: We deploy a rigid RIS with $N$ elements under identical simulation settings. The rigid RIS only performs phase shift optimization and beamforming optimization, without the additional flexibility of surface shape morphing. 
 \item \textbf{Random surface shape}: We assume that each element in the FIM surface shape vector is generated uniformly and independently from $[-d_{\text{max}},d_{\text{max}}]$. Only the active beamforming and phase shift optimization are considered.
 \item \textbf{Random phase shift}: We assume that the phase of each FIM element is independently and uniformly produced from $[0, 2\pi]$, when applying the optimal beamforming and surface shape.
 \item \textbf{Without FIM}: Let $N=0$ and only optimize the active beamforming.
\end{itemize}

\subsection{Performance Analysis}
In Fig. 4, we evaluate the convergence behavior of the BCD algorithms when $P_t = 30$ dBm and $N=30$. Fig. 4 shows that the proposed algorithm converges rapidly, with a convergence rate comparable to that designed for the RIS-based benchmark. This indicates that morphing the FIM surface shape does not significantly increase the overall algorithmic complexity, which aligns with our complexity analysis in Section \ref{subsec:overall}, where we showed that the computational complexity of the proposed FIM optimization method remains linear with respect to $N$. Furthermore, it can be observed that morphing the surface shape of FIM significantly enhances system performance compared to other baseline schemes. In particular, compared to the rigid RIS-based scheme, the proposed FIM-assisted system achieves an approximately 33\% improvement in terms of WSR, highlighting the benefits of leveraging FIM's additional flexibility for reshaping the wireless environment.

Fig. 5 illustrates the performance of proposed algorithm with imperfect CSI when $P_t = 30$ dBm and $N=30$. Specifically, Fig. 5(a) shows that imperfect CSI leads to performance degradation. Nevertheless, the WSR of the FIM-assisted system remains higher than that of the RIS-assisted counterpart. Moreover, when the MSE $\varsigma \leq 0.8$, the degradation in WSR does not exceed 30\%, which demonstrates the robustness of the proposed algorithm. Fig. 5(b) shows that the system WSR decreases as the MSE increases, which is consistent with expectations. 

Fig.~6(a) shows the convergence behavior of the algorithm under different SNR levels under $N=30$. It can be observed that the algorithm converges rapidly across various SNR levels, and a higher SNR leads to a larger WSR, which aligns with expectations. Fig.~6(b) further illustrates the impact of SNR on the number of iterations and the runtime. As shown, a higher SNR reduces the number of iterations required for convergence as well as the average runtime.\footnote{All runtime results were obtained using a single-threaded serial implementation without extensive code optimization to ensure a fair comparison with all baseline methods.}

\begin{figure}[t] 
\centering %
\includegraphics[width=0.42\textwidth]{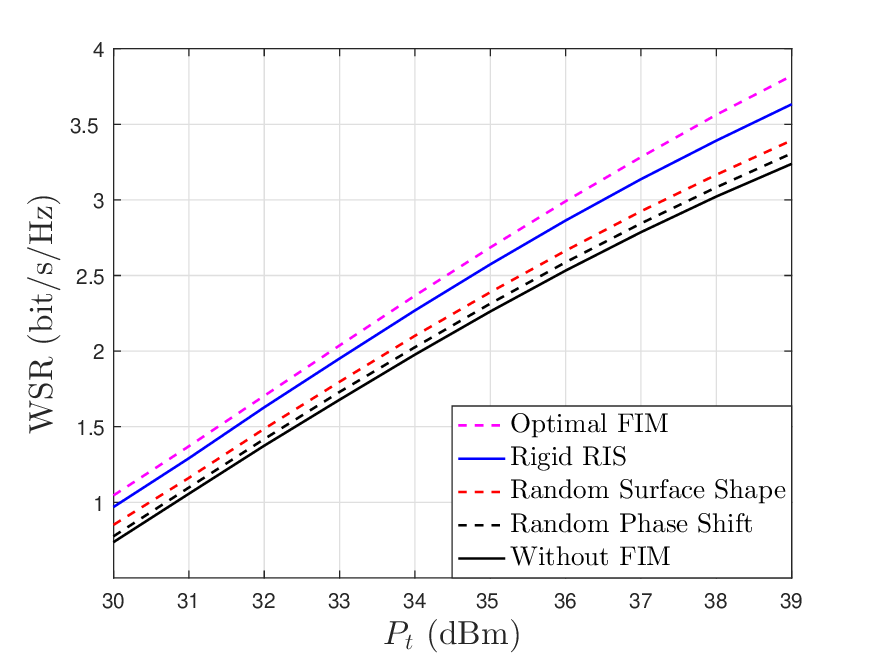}
\caption{WSR versus the transmit power $P_t$ when considering $N=30$.}
\label{fig_1}
\vspace{-0.5cm}
\end{figure}
\begin{figure}[t] 
\centering %
\includegraphics[width=0.42\textwidth]{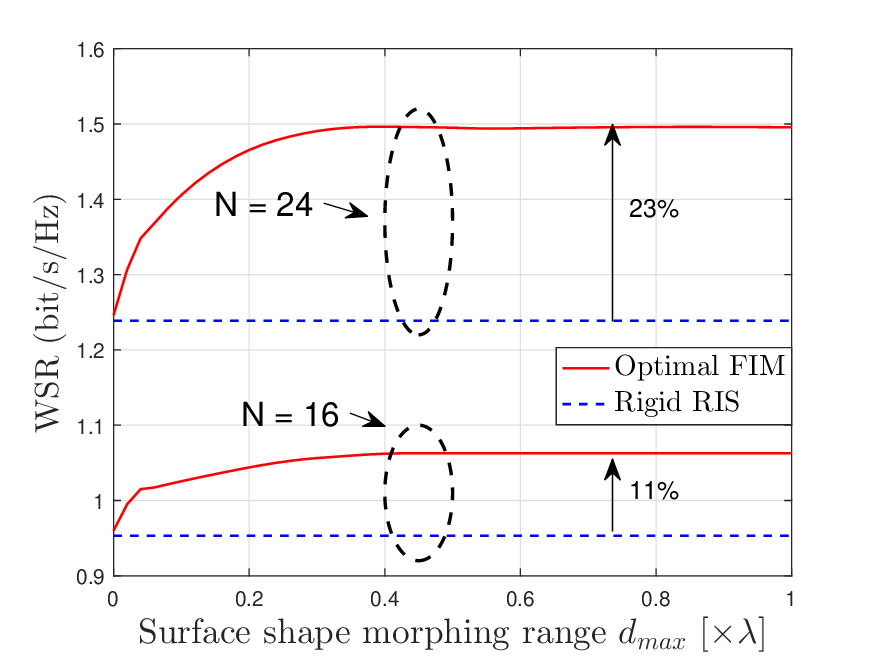}
\caption{WSR versus the morphing range $d_{\text{max}}$ under different values of $N$.}
\label{fig_1}
\vspace{-0.5cm}
\end{figure}

Fig. 7 illustrates the relationship between the transmit power $P_t$ and the WSR, when $N = 30$. It is observed that the WSR achieved by different schemes increases as $P_t$ increases. As expected, if the phase shift matrix or the FIM surface shape are randomly configured, the performance gain brought by deploying FIM is nearly negligible. Notably, even with phase shift optimization, the system performance under a randomly morphed surface shape is observed to be inferior to that of a rigid RIS. This is because a random surface shape configuration may destructively combine multipath channels, leading to performance degradation. In contrast, an optimized FIM surface shape constructively aligns multipath components, allowing the FIM-assisted system to consistently outperform the rigid RIS-assisted counterpart as the transmit power increases. 

Fig. 8 depicts the impact of the surface shape morphing range $d_{\text{max}}$ on the WSR under two different configurations: $i)$ $N = 16$ and $ii)$ $N = 24$. First, as $d_{\text{max}}$ increases, the WSR improves correspondingly. This is because the FIM exhibits a superior capability for surface shape morphing, thereby enhancing system performance. Second, a diminishing return effect is observed. The root of this phenomenon is the periodic dependence of the objective function on the surface shape, which is a direct consequence of the steering vector's complex exponential form. As a consequence, once the first global optimum is attained, further enlargement of $d_{\text{max}}$ yields no additional performance improvement, resulting in a saturation of the curve. From Fig. 8, we can observe that the globally optimal surface shape is achieved through sub-wavelength morphing, which makes the deployment of FIM possible. Third, comparing WSR under different values of $N$, we observe that a larger value of $N$ results in a higher WSR, which aligns with expectations. Finally, when comparing the performance achieved by FIM and the rigid RIS, we find that the performance gap between the two schemes increases as $N$ increases. Specifically, when $N = 16$, the WSR difference between FIM and RIS is 0.11 bit/s/Hz (Relative increase: 11\%), whereas this gap expands to 0.26 bit/s/Hz (Relative increase: 24\%) when $N = 24$. This is because as $N$ increases, the FIM gains an increased dimensionality to morph its surface shape. Therefore, the performance improvement of FIM over RIS cannot be easily offset by merely increasing the number of elements, and such hardware cost rises continuously as the target performance metric increases.

\begin{figure}[t] 
\centering %
\includegraphics[width=0.42\textwidth]{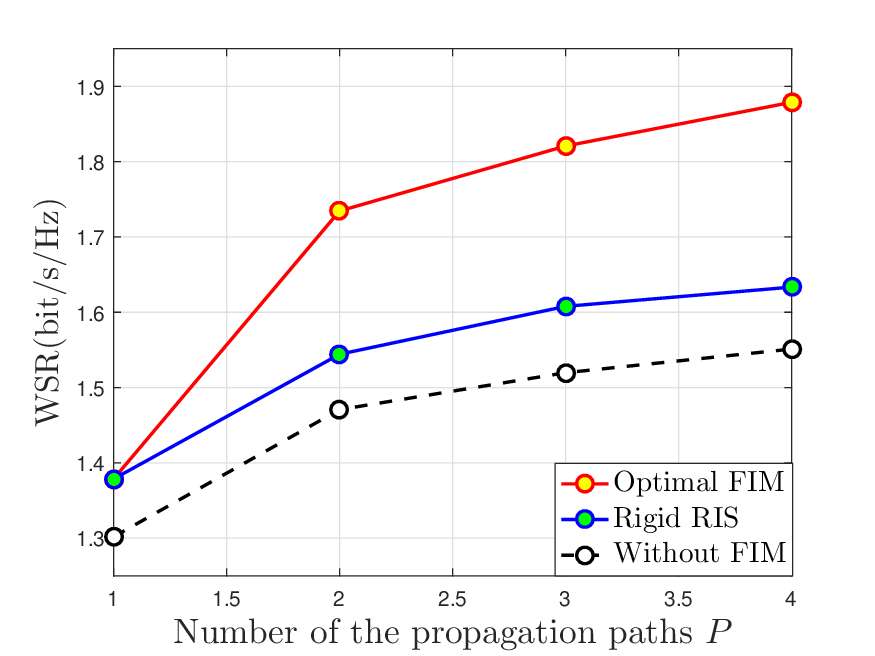}
\caption{WSR versus the number of propagation paths $P$.}
\label{fig_1}
\vspace{-0.4cm}
\end{figure}
\begin{figure}[!t]
\centering
\subfloat[]{\includegraphics[width=0.25\textwidth]{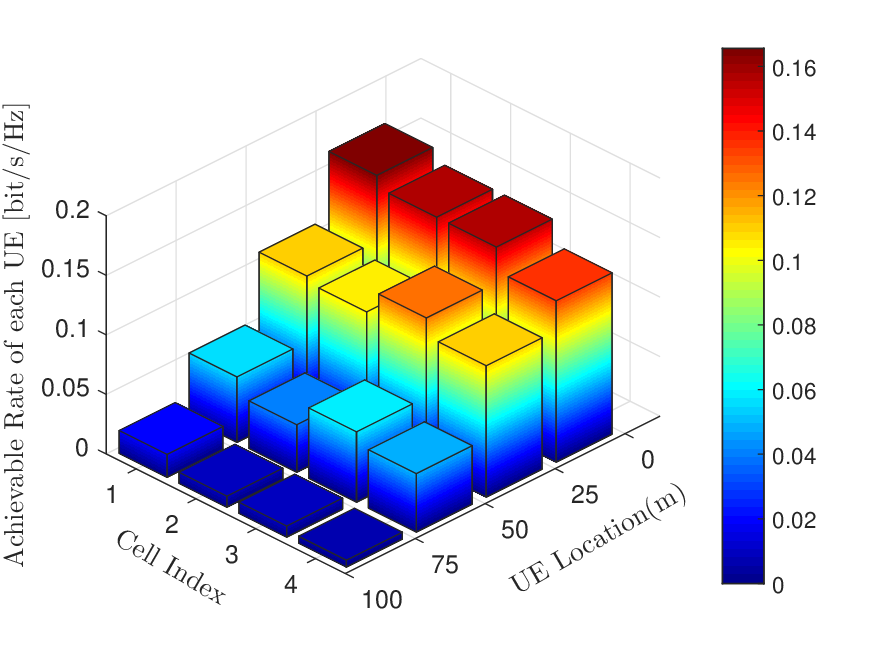}}
\subfloat[]{\includegraphics[width=0.25\textwidth]{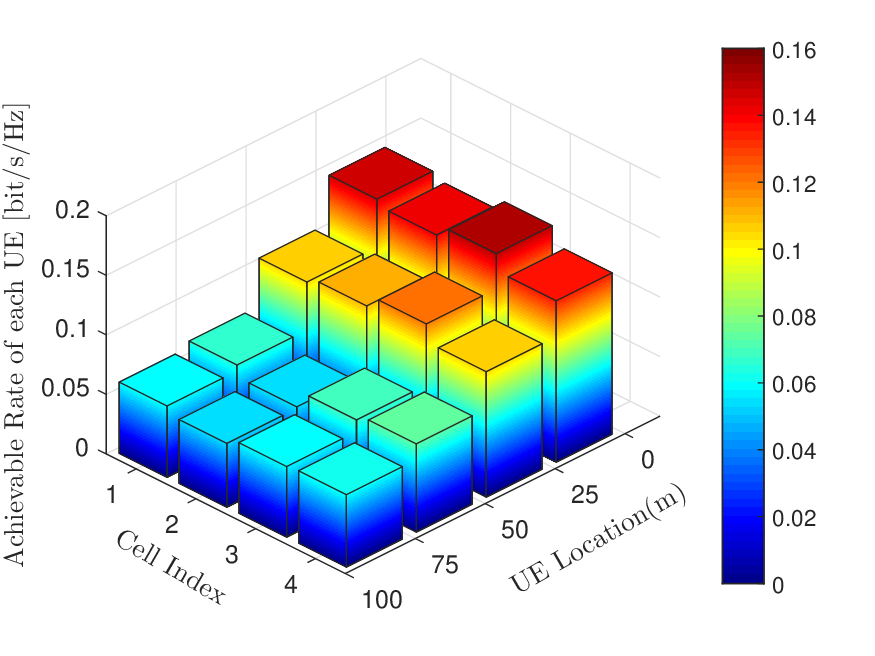}}
\caption{(a) Achievable rate of each UE aided by RIS; (b) Achievable rate of each UE aided by FIM.}
\label{fig_3}
\vspace{-0.5cm}
\end{figure}
In Fig. 9, the overall WSR of the system under various propagation paths in the FIM-UE link is evaluated, when $P_t = 30$ dBm and $N=30$. The results reveal that a growing number of multipath components leads to a marked improvement in the WSR performance gain of FIM, far outperforming the RIS-based scheme. Moreover, the performance gap between the two schemes widens by increasing the number of paths. This demonstrates the effectiveness of FIM in alleviating the detrimental effects of multipath fading.

In Fig. 10, we investigate the effect of FIM by examining the achievable rates of different UEs. Specifically, we consider a system with 4 cells $(L=4)$, each serving 4 UEs $(K=4)$, where $N=30, P_t=30$ dBm. Fig. 10(a) reveals that, even with the assistance of RIS, users closer to the cell edge experience significantly lower achievable rates due to severe inter-cell interference. Specifically, users located at 75–100 m away from the cell center maintain an achievable rate about 0.01 bit/s/Hz, severely degrading the system's overall performance. After deploying an FIM, the rates of users shown in Fig. 10(b) are notably improved, achieving about 0.05 bit/s/Hz. This confirms that FIM effectively mitigates inter-cell interference and enhances the QoS for edge users. Furthermore, if the weight factor for cell-edge users is deliberately increased, the system's communication performance can be further improved.

\begin{figure}[t] 
\centering %
\includegraphics[width=0.44\textwidth]{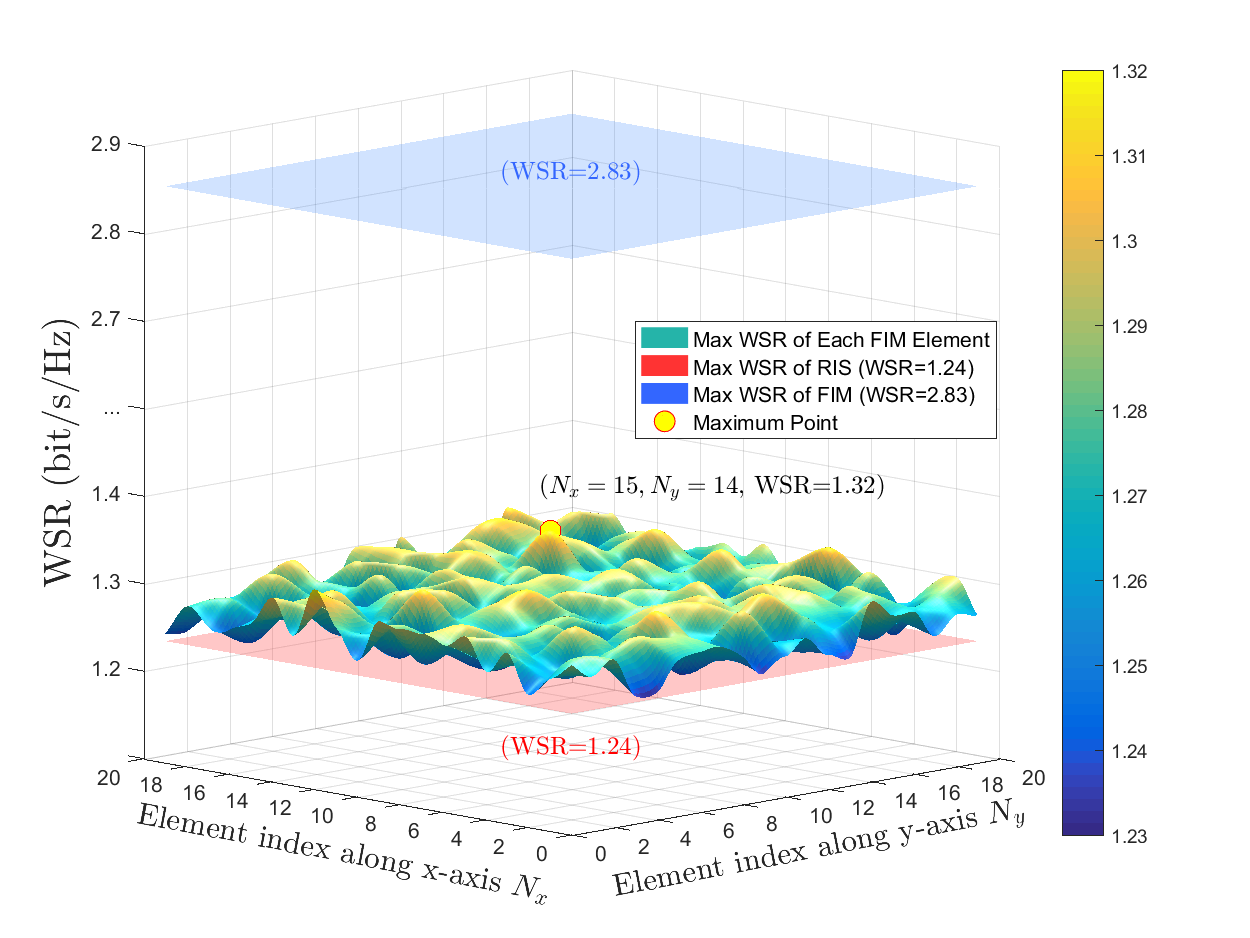}
\caption{Comparison between the single-element displacement optimization and overall surface shape morphing.} 
\label{fig_1}
\vspace{-0.6cm}
\end{figure}

To further illustrate the performance gain obtained by morphing the FIM surface shape, Fig. 11 compares the WSR performance under three configurations: $i)$ RIS-assisted systems, $ii)$ FIM-assisted systems employing only a single element's displacement is optimized, and $iii)$ FIM-assisted systems using the proposed algorithm for morphing the surface shape. The simulation considers a square metasurface with $N = 400$ elements and the transmit power of $P_t = 30$ dBm. 
It can be observed that the RIS-assisted baseline achieves a WSR of 1.24 bit/s/Hz. In contrast, optimizing the displacement of any individual element in the FIM yields a WSR higher than this benchmark. 
Specifically, the element located at coordinates $(N_x=15$, $N_y=14)$ achieves the highest WSR of 1.32 bit/s/Hz when adopting the single-element optimization method. In contrast, the proposed joint optimization approach achieves a significantly higher WSR of 2.83 bit/s/Hz. These findings reveal two key conclusions. Firstly, FIM demonstrates significantly greater potential than RIS in enhancing communication performance. Even optimizing a single or a small number of deformable units can outperform RIS, indicating that the performance achieved by RIS serves as a lower bound for characterizing FIM's capabilities. Secondly, joint optimization is crucial for realizing the full benefits of FIM, highlighting the necessity of optimizing the holistic FIM surface shape.

\section{Conclusion}
\label{sec:Conclusion}
In this paper, we explored the deployment of an FIM at the cell boundary to improve the WSR of a multicell MU-MISO system. To achieve this goal, we jointly optimized the beamforming vectors, the phase shift matrix at the FIM, and the surface shape of the FIM, subject to constraints on the transmit power budget, unit-modulus reflection coefficients, and surface shape morphing range To tackle the non-convex and highly coupled optimization problem, we employed an alternating optimization framework. The WMMSE method was applied to simplify the objective function and yield a closed-form solution for beamforming, while the BCD algorithm was used to decouple the variables and optimize them iteratively. Specifically, we utilized the RCG method for phase shift optimization, and the PGD algorithm for surface shape optimization. Our simulation results confirm the effectiveness of the BCD algorithms and showcase the fast convergence of the optimization method. Furthermore, FIM significantly outperformed conventional RIS in terms of WSR. In particular, optimal surface shaping was achieved within a small morphing range, ensuring the feasibility of the physical implementation of FIM. In future research, addressing the dynamic misalignment caused by the temporal mismatch between phase shifting and shape morphing will be a key focus of discussion. Furthermore, the power consumption modeling and energy efficiency of FIMs also constitute an important research direction.

\appendix
This appendix provides the detailed derivation steps for transforming the original WMMSE objective function (20a) into the respective subproblems. Firstly, we can further expand the MSE parameter $e_{l,k}$ based on (16), resulting in 
% 左栏底部
\begin{align}
    &e_{l,k} = (u^*_{l,k} \mathbf{h}_{l,l,k}\mathbf{w}_{l,k} - 1) (u^*_{l,k} \mathbf{h}_{l,l,k}\mathbf{w}_{l,k} - 1)^* \notag\\
    &+ \sum_{\hat{k}=1,\hat{k}\neq k}^K \left|u_{l,k}\right|^2 \mathbf{h}_{l,l,k} \mathbf{w}_{l,\hat{k}} \mathbf{w}^\mathrm{H}_{l,\hat{k}} \mathbf{h}^\mathrm{H}_{l,l,k}\notag\\
    &+ \sum_{\hat{l}=1,\hat{l}\neq l}^L \sum_{\hat{k}=1}^K \left|u_{l,k}\right|^2\mathbf{h}_{\hat{l},l,k} \mathbf{w}_{\hat{l},\hat{k}} \mathbf{w}^\mathrm{H}_{\hat{l},\hat{k}} \mathbf{h}^\mathrm{H}_{\hat{l},l,k}+ \left|u_{l,k}\right|^2\sigma_{l,k}^2  \tag{72}
\end{align}
\begin{align}
&= |u_{l,k}|^2 \mathbf{h}_{l,l,k} \mathbf{w}_{l,k} \mathbf{w}_{l,k}^\mathrm{H} \mathbf{h}_{l,l,k}^\mathrm{H} - 2\operatorname{Re}(u_{l,k}^* \mathbf{h}_{l,l,k} \mathbf{w}_{l,k}) + 1 \notag\\
&+ |u_{l,k}|^2 \sum_{\hat{k}\neq k}^K \mathbf{h}_{l,l,k} \mathbf{w}_{l,\hat{k}} \mathbf{w}_{l,\hat{k}}^\mathrm{H} \mathbf{h}_{l,l,k}^\mathrm{H} + |u_{l,k}|^2 \sigma_{l,k}^2 \notag\\
&+ |u_{l,k}|^2 \sum_{\hat{l}\neq l}^L \sum_{\hat{k}=1}^K \mathbf{h}_{\hat{l},l,k} \mathbf{w}_{\hat{l},\hat{k}} \mathbf{w}_{\hat{l},\hat{k}}^\mathrm{H} \mathbf{h}_{\hat{l},l,k}^\mathrm{H} \tag{73}\\
&= |u_{l,k}|^2 \sum_{\hat{l}=1}^L \sum_{\hat{k}=1}^K \mathbf{h}_{\hat{l},l,k} \mathbf{w}_{\hat{l},\hat{k}} \mathbf{w}_{\hat{l},\hat{k}}^\mathrm{H} \mathbf{h}_{\hat{l},l,k}^\mathrm{H} \notag\\
& - 2\operatorname{Re}(u_{l,k}^* \mathbf{h}_{l,l,k} \mathbf{w}_{l,k}) + |u_{l,k}|^2 \sigma_{l,k}^2 + 1 \tag{74}
\end{align}

Next, by substituting (74) and performing matrix operations, we can obtain each simplified subproblem. Taking the optimization of beamforming as an example, we ignore the additive terms unrelated to $\mathbf{w}_{l,k}$ and substitute the above expression. After consolidating the multiplicative terms independent of $\mathbf{w}_{l,k}$, we can obtain the objective function as (24a).
\bibliographystyle{IEEEtran}
\bibliography{IEEEabrv,reference}

\begin{thebibliography}{10}
\providecommand{\url}[1]{#1}
\csname url@samestyle\endcsname
\providecommand{\newblock}{\relax}
\providecommand{\bibinfo}[2]{#2}
\providecommand{\BIBentrySTDinterwordspacing}{\spaceskip=0pt\relax}
\providecommand{\BIBentryALTinterwordstretchfactor}{4}
\providecommand{\BIBentryALTinterwordspacing}{\spaceskip=\fontdimen2\font plus
\BIBentryALTinterwordstretchfactor\fontdimen3\font minus \fontdimen4\font\relax}
\providecommand{\BIBforeignlanguage}[2]{{%
\expandafter\ifx\csname l@#1\endcsname\relax
\typeout{** WARNING: IEEEtran.bst: No hyphenation pattern has been}%
\typeout{** loaded for the language `#1'. Using the pattern for}%
\typeout{** the default language instead.}%
\else
\language=\csname l@#1\endcsname
\fi
#2}}
\providecommand{\BIBdecl}{\relax}
\BIBdecl

\bibitem{11431856}
H.~Hu, J.~An, L.~Gan, A.~Nallanathan, and N.~Al-Dhahir, ``Weighted sum-rate maximization for flexible intelligent metasurface aided multicell systems,'' in \emph{GLOBECOM 2025 - 2025 IEEE Global Communications Conference}, 2025, pp. 3618--3623.

\bibitem{ref1}
H.~Tataria, M.~Shafi, A.~F. Molisch, M.~Dohler, H.~Sj{\"o}land, and F.~Tufvesson, ``6{G} wireless systems: Vision, requirements, challenges, insights, and opportunities,'' \emph{Proc. IEEE}, vol. 109, no.~7, pp. 1166--1199, Jul. 2021.

\bibitem{ref2}
M.~Giordani, M.~Polese, M.~Mezzavilla, S.~Rangan, and M.~Zorzi, ``Toward 6{G} networks: Use cases and technologies,'' \emph{IEEE Commun. Mag.}, vol.~58, no.~3, pp. 55--61, Mar. 2020.

\bibitem{ref3}
K.~B. Letaief, W.~Chen, Y.~Shi, J.~Zhang, and Y.-J.~A. Zhang, ``The roadmap to 6{G}: {AI} empowered wireless networks,'' \emph{IEEE Commun. Mag.}, vol.~57, no.~8, pp. 84--90, Aug. 2019.

\bibitem{ref4}
M.~D. Renzo, A.~Zappone, M.~Debbah, M.-S. Alouini, C.~Yuen, J.~de~Rosny, and S.~Tretyakov, ``Smart radio environments empowered by reconfigurable intelligent surfaces: How it works, state of research, and the road ahead,'' \emph{IEEE J. Sel. Areas Commun.}, vol.~38, no.~11, pp. 2450--2525, Nov. 2020.

\bibitem{ref5}
T.~J. Cui, M.~Q. Qi, X.~Wan, J.~Zhao, and Q.~Cheng, ``Coding metamaterials, digital metamaterials and programmable metamaterials,'' \emph{Light Sci. Appl.}, vol.~3, no.~10, pp. e218--e218, Oct. 2014.

\bibitem{ref6}
Y.~Liu, X.~Liu, X.~Mu, T.~Hou, J.~Xu, M.~D. Renzo, and N.~Al-Dhahir, ``Reconfigurable intelligent surfaces: Principles and opportunities,'' \emph{IEEE Commun. Surv. Tutorials}, vol.~23, no.~3, pp. 1546--1577, Third Quarter 2021.

\bibitem{ref7}
J.~An, C.~Xu, Q.~Wu, D.~W.~K. Ng, M.~D. Renzo, C.~Yuen, and L.~Hanzo, ``Codebook-based solutions for reconfigurable intelligent surfaces and their open challenges,'' \emph{IEEE Wireless Commun.}, vol.~31, no.~2, pp. 134--141, Apr. 2024.

\bibitem{ref8}
Z.~Chen, G.~Chen, J.~Tang, S.~Zhang, D.~K. So, O.~A. Dobre, K.~K. Wong, and J.~Chambers, ``Reconfigurable-intelligent-surface-assisted {B}5{G}/6{G} wireless communications: Challenges, solution, and future opportunities,'' \emph{IEEE Commun. Mag.}, vol.~61, no.~1, pp. 16--22, Jan. 2023.

\bibitem{ref9}
L.~Dai, B.~Wang, M.~Wang, X.~Yang, J.~Tan, S.~Bi, S.~Xu, F.~Yang, Z.~Chen, M.~D. Renzo, C.-B. Chae, and L.~Hanzo, ``Reconfigurable intelligent surface-based wireless communications: Antenna design, prototyping, and experimental results,'' \emph{IEEE Access}, vol.~8, pp. 45\,913--45\,923, 2020.

\bibitem{ref10}
C.~Huang, A.~Zappone, G.~C. Alexandropoulos, M.~Debbah, and C.~Yuen, ``Reconfigurable intelligent surfaces for energy efficiency in wireless communication,'' \emph{IEEE Trans. Wireless Commun.}, vol.~18, no.~8, pp. 4157--4170, Aug. 2019.

\bibitem{ref11}
\BIBentryALTinterwordspacing
H.~Hu, J.~An, L.~Gan, and C.~Yuen, ``Performance analysis of {RIS}-aided high-mobility wireless systems,'' 2025. [Online]. Available: \url{https://arxiv.org/abs/2508.15375}
\BIBentrySTDinterwordspacing

\bibitem{ref12}
J.~An, C.~Xu, L.~Gan, and L.~Hanzo, ``Low-complexity channel estimation and passive beamforming for {RIS}-assisted {MIMO} systems relying on discrete phase shifts,'' \emph{IEEE Trans. Commun.}, vol.~70, no.~2, pp. 1245--1260, Feb. 2022.

\bibitem{ref13}
A.~Shojaeifard, K.-K. Wong, K.-F. Tong, Z.~Chu, A.~Mourad, A.~Haghighat, I.~Hemadeh, N.~T. Nguyen, V.~Tapio, and M.~Juntti, ``{MIMO} evolution beyond {5G} through reconfigurable intelligent surfaces and fluid antenna systems,'' \emph{Proc. IEEE}, vol. 110, no.~9, pp. 1244--1265, Sept. 2022.

\bibitem{ref14}
Q.~Wu and R.~Zhang, ``Beamforming optimization for wireless network aided by intelligent reflecting surface with discrete phase shifts,'' \emph{IEEE Trans. Commun.}, vol.~68, no.~3, pp. 1838--1851, Mar. 2020.

\bibitem{ref15}
S.~Hu, F.~Rusek, and O.~Edfors, ``Beyond massive {MIMO}: The potential of data transmission with large intelligent surfaces,'' \emph{IEEE Trans. Signal Process.}, vol.~66, no.~10, pp. 2746--2758, May 2018.

\bibitem{ref16}
E.~Bj\"{o}rnson, O.~\"{O}., and E.~G. Larsson, ``Intelligent reflecting surface vs. decode-and-forward: How large surfaces are needed to beat relaying?'' \emph{IEEE Wireless Commun. Lett.}, vol.~9, no.~2, pp. 244--248, Feb. 2020.

\bibitem{ref17}
K.~Zhi, C.~Pan, H.~Ren, and K.~Wang, ``Statistical {CSI}-based design for reconfigurable intelligent surface-aided massive {MIMO} systems with direct links,'' \emph{IEEE Wireless Commun. Lett.}, vol.~10, no.~5, pp. 1128--1132, May 2021.

\bibitem{ref18}
G.~T. de~Ara{\'u}jo and A.~L.~F. de~Almeida, ``Parafac-based channel estimation for intelligent reflective surface assisted {MIMO} system,'' in \emph{Proc. IEEE Sens. Array Multichannel Signal Process. Workshop (SAM)}, Hangzhou, China, 2020, pp. 1--5.

\bibitem{ref19}
J.~Chen, Y.-C. Liang, H.~V. Cheng, and W.~Yu, ``Channel estimation for reconfigurable intelligent surface aided multi-user mmwave {MIMO} systems,'' \emph{IEEE Trans. Wireless Commun.}, vol.~22, no.~10, pp. 6853--6869, Oct. 2023.

\bibitem{ref20}
E.~Shi, J.~Zhang, H.~Du, B.~Ai, C.~Yuen, D.~Niyato, K.~B. Letaief, and X.~Shen, ``{RIS}-aided cell-free massive {MIMO} systems for 6{G}: Fundamentals, system design, and applications,'' \emph{Proc. IEEE}, vol. 112, no.~4, pp. 331--364, Apr. 2024.

\bibitem{ref21}
C.~Pan, H.~Ren, K.~Wang, W.~Xu, M.~Elkashlan, A.~Nallanathan, and L.~Hanzo, ``Multicell {MIMO} communications relying on intelligent reflecting surfaces,'' \emph{IEEE Trans. Wireless Commun.}, vol.~19, no.~8, pp. 5218--5233, Aug. 2020.

\bibitem{ref22}
Y.~Zhang, B.~Di, H.~Zhang, J.~Lin, Y.~Li, and L.~Song, ``Reconfigurable intelligent surface aided cell-free {MIMO} communications,'' \emph{IEEE Wireless Commun. Lett.}, vol.~10, no.~4, pp. 775--779, Apr. 2021.

\bibitem{ref23}
X.~Yu, D.~Xu, and R.~Schober, ``{MISO} wireless communication systems via intelligent reflecting surfacess: Joint active and passive beamforming optimization,'' \emph{IEEE Trans. Commun.}, vol.~72, no.~1, pp. 408--424, Jan. 2024.

\bibitem{ref26}
X.~Wang, Z.~Fei, J.~Guo, Z.~Zheng, and B.~Li, ``{RIS}-assisted spectrum sharing between {MIMO} radar and {MU}-{MISO} communication systems,'' \emph{IEEE Wireless Commun. Lett.}, vol.~10, no.~3, pp. 594--598, Mar. 2021.

\bibitem{ref27}
W.~Tang, J.~Y. Dai, M.~Z. Chen, K.-K. Wong, X.~Li, X.~Zhao, S.~Jin, Q.~Cheng, and T.~J. Cui, ``{MIMO} transmission through reconfigurable intelligent surface: System design, analysis, and implementation,'' \emph{IEEE J. Sel. Areas Commun.}, vol.~38, no.~11, pp. 2683--2699, Nov. 2020.

\bibitem{ref28}
X.~Yu, J.~Shen, J.~Zhang, and K.~B. Letaief, ``Alternating minimization algorithms for hybrid precoding in millimeter wave {MIMO} systems,'' \emph{IEEE J. Sel. Topics Signal Process.}, vol.~10, no.~3, pp. 485--500, 2016.

\bibitem{ref29}
S.~Buzzi, C.~D’Andrea, A.~Zappone, M.~Fresia, Y.-P. Zhang, and S.~Feng, ``{RIS} configuration, beamformer design, and power control in single-cell and multi-cell wireless networks,'' \emph{IEEE Trans. Cogn. Commun. Netw.}, vol.~7, no.~2, pp. 398--411, June 2021.

\bibitem{ref30}
H.~Gao, X.~Yang, N.~Chen, S.~Chen, Y.~Yang, and C.~Yuen, ``Robust beamforming for reconfigurable intelligent surface-assisted multi-cell downlink transmissions,'' \emph{IEEE Trans. Veh. Technol.}, vol.~73, no.~5, pp. 6910--6922, May 2024.

\bibitem{ref31}
M.~Cui, G.~Zhang, and R.~Zhang, ``Secure wireless communication via intelligent reflecting surface,'' \emph{IEEE Wireless Commun. Lett.}, vol.~8, no.~5, pp. 1410--1414, Oct. 2019.

\bibitem{ref32}
H.~Shen, W.~Xu, W.~Xu, S.~Gong, Z.~He, and C.~Zhao, ``Secrecy rate maximization for intelligent reflecting surface assisted multiantenna communications,'' \emph{IEEE Commun. Lett.}, vol.~23, no.~9, pp. 1488--1492, Sept. 2019.

\bibitem{ref34}
J.~An, C.~Xu, D.~W.~K. Ng, G.~C. Alexandropoulos, C.~Huang, C.~Yuen, and L.~Hanzo, ``Stacked intelligent metasurfaces for efficient holographic {MIMO} communications in 6{G},'' \emph{IEEE J. Sel. Areas Commun.}, vol.~41, no.~8, pp. 2380--2396, Aug. 2023.

\bibitem{ref36}
E.~Shi, J.~Zhang, J.~An, G.~Zhang, Z.~Liu, C.~Yuen, and B.~Ai, ``Joint {AP-UE} association and precoding for {SIM}-aided cell-free massive {MIMO} systems,'' \emph{IEEE Trans. Wireless Commun.}, vol.~24, no.~6, pp. 5352--5367, Jun. 2025.

\bibitem{ref37}
E.~Shi, J.~Zhang, Y.~Zhu, J.~An, C.~Yuen, and B.~Ai, ``Uplink performance of stacked intelligent metasurface-enhanced cell-free massive {MIMO} systems,'' \emph{IEEE Trans. Wireless Commun.}, vol.~24, no.~5, pp. 3731--3746, May 2025.

\bibitem{ref38}
H.~Liu, J.~An, X.~Jia, L.~Gan, G.~K. Karagiannidis, B.~Clerckx, M.~Bennis, M.~Debbah, and T.~J. Cui, ``Stacked intelligent metasurfaces for wireless communications: Applications and challenges,'' \emph{IEEE Wireless Commun.}, vol.~32, no.~4, pp. 46--53, Aug. 2025.

\bibitem{ref33}
J.~An, C.~Yuen, C.~Xu, H.~Li, D.~W.~K. Ng, M.~D. Renzo, M.~Debbah, and L.~Hanzo, ``Stacked intelligent metasurface-aided {MIMO} transceiver design,'' \emph{IEEE Wireless Commun.}, vol.~31, no.~4, pp. 123--131, Aug. 2024.

\bibitem{ref35}
J.~An, M.~D. Renzo, M.~Debbah, H.~V. Poor, and C.~Yuen, ``Stacked intelligent metasurfaces for multiuser downlink beamforming in the wave domain,'' \emph{IEEE Trans. Wireless Commun.}, vol.~24, no.~7, pp. 5525--5538, Jul. 2025.

\bibitem{ref39}
D.~Tse and P.~Viswanath, ``Fundamentals of wireless communication,'' 2005.

\bibitem{ref40}
X.~Ni, H.~Luan, J.~T. Kim \emph{et~al.}, ``Soft shape-programmable surfaces by fast electromagnetic actuation of liquid metal networks,'' \emph{Nat. Commun.}, vol.~13, p. 5576, Sep. 2022.

\bibitem{ref41}
Y.~Bai, H.~Wang, Y.~Xue, Y.~Pan, J.-T. Kim, X.~Ni, T.-L. Liu, Y.~Yang, M.~Han, and Y.~e.~a. Huang, ``A dynamically reprogrammable surface with self-evolving shape morphing,'' \emph{Nature}, vol. 609, no. 7928, pp. 701--708, Jul. 2022.

\bibitem{ref42}
D.~Niu, W.~Jiang, D.~Li, G.~Ye, F.~Luo, and H.~Liu, ``Reconfigurable shape-morphing flexible surfaces realized by individually addressable photoactuator arrays,'' \emph{Smart Mater. Struct.}, vol.~30, no.~12, p. 125032, Dec. 2021.

\bibitem{ref43}
Y.~Zhou, S.~Wang, J.~Yin, J.~Wang, F.~Manshaii, X.~Xiao, T.~Zhang, H.~Bao, S.~Jiang, and J.~Chen, ``Flexible metasurfaces for multifunctional interfaces,'' \emph{ACS Nano}, vol.~18, no.~4, pp. 2685--2707, Jan. 2024.

\bibitem{ref24}
Q.~Wu and R.~Zhang, ``Intelligent reflecting surface enhanced wireless network via joint active and passive beamforming,'' \emph{IEEE Trans. Wireless Commun.}, vol.~18, no.~11, pp. 5394--5409, Nov. 2019.

\bibitem{ref25}
H.~Guo, Y.-C. Liang, J.~Chen, and E.~G. Larsson, ``Weighted sum-rate maximization for reconfigurable intelligent surface aided wireless networks,'' \emph{IEEE Trans. Wireless Commun.}, vol.~19, no.~5, pp. 3064--3076, May 2020.

\bibitem{ref44}
J.~An, Z.~Han, D.~Niyato, M.~Debbah, C.~Yuen, and L.~Hanzo, ``Flexible intelligent metasurfaces for enhancing {MIMO} communications,'' \emph{IEEE Trans. Commun.}, pp. 1--1, Mar. 2025.

\bibitem{ref45}
J.~An, C.~Yuen, M.~{Di Renzo}, M.~Debbah, H.~V. Poor, and L.~Hanzo, ``Flexible intelligent metasurfaces for downlink multiuser {MISO} communications,'' \emph{IEEE Trans. Wireless Commun.}, vol.~24, no.~4, pp. 2940--2955, Apr. 2025.

\bibitem{ref46}
\BIBentryALTinterwordspacing
K.~R.~R. Ranasinghe, J.~An, I.~A.~M. Sandoval, H.~S. Rou, G.~T.~F. de~Abreu, C.~Yuen, and M.~Debbah, ``Flexible intelligent metasurfaces in high-mobility {MIMO} integrated sensing and communications,'' 2025. [Online]. Available: \url{https://arxiv.org/abs/2507.18793}
\BIBentrySTDinterwordspacing

\bibitem{ref47}
Z.~Teng, J.~An, L.~Gan, N.~Al-Dhahir, and Z.~Han, ``Flexible intelligent metasurface for enhancing multi-target wireless sensing,'' \emph{IEEE Trans. Veh. Technol.}, pp. 1--6, 2025, early Access Article.

\bibitem{ref48}
J.~An, M.~Debbah, T.~J. Cui, Z.~N. Chen, and C.~Yuen, ``Emerging technologies in intelligent metasurfaces: Shaping the future of wireless communications,'' \emph{IEEE Trans. Antennas Propag.}, pp. 1--1, 2025, early Access Article.

\bibitem{ref49}
\BIBentryALTinterwordspacing
S.~Yang, Z.~Wan, B.~Ning, W.~Mei, J.~An, Y.~C. Eldar, and C.~Yuen, ``Flexible intelligent metasurface-aided wireless communications: Architecture and performance,'' 2025. [Online]. Available: \url{https://arxiv.org/abs/2503.11112}
\BIBentrySTDinterwordspacing

\bibitem{ref53}
H.~Hu, J.~An, L.~Gan, and N.~Al-Dhahir, ``Flexible intelligent metasurface for reconfiguring radio environments,'' \emph{IEEE Trans. Veh. Technol}, pp. 1--6, Oct. 2025.

\bibitem{ref50}
Q.~Shi, M.~Razaviyayn, Z.~Q. Luo, and C.~He, ``An iteratively weighted mmse approach to distributed sum-utility maximization for a {MIMO} interfering broadcast channel,'' \emph{IEEE Trans. Signal Process.}, vol.~59, no.~9, pp. 4331--4340, Sept. 2011.

\bibitem{ref51}
K.~Alhujaili, V.~Monga, and M.~Rangaswamy, ``Transmit {MIMO} radar beampattern design via optimization on the complex circle manifold,'' \emph{IEEE Trans. Signal Process.}, vol.~67, no.~13, pp. 3561--3575, Jul. 2019.

\bibitem{ref54}
C.~Pan, H.~Ren, K.~Wang, J.~Xu, W.~Li, Y.~He, and Y.-C. Zhang, ``An overview of signal processing techniques for ris/irs-aided wireless systems,'' \emph{IEEE J. Sel. Top. Signal Process.}, vol.~16, no.~5, pp. 883--917, Aug. 2022.

\bibitem{ref52}
\BIBentryALTinterwordspacing
A.~Jiménez-Sáez, A.~Asadi, R.~Neuder, M.~Delbari, and V.~Jamali, ``Reconfigurable intelligent surfaces with liquid crystal technology: A hardware design and communication perspective,'' 2023. [Online]. Available: \url{https://arxiv.org/abs/2308.03065}
\BIBentrySTDinterwordspacing

\end{thebibliography}

\begin{IEEEbiography}
[{\includegraphics[width=1in,clip,keepaspectratio]{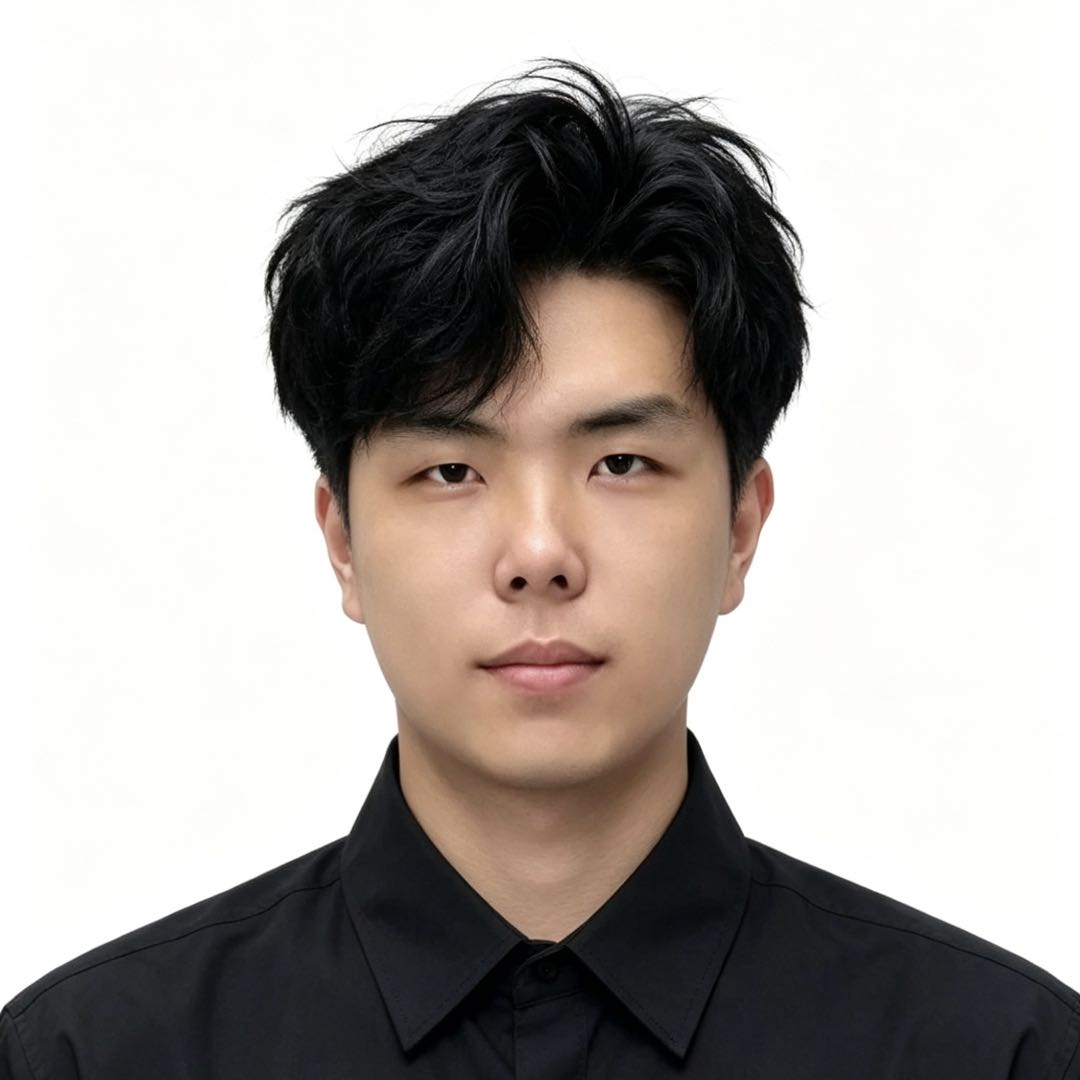}}]
{Hanwen Hu} (Member, IEEE) received the B.S. degree in Information Engineering from Northwestern Polytechnical University, Xi'an, China, in 2023. He is currently pursuing the Ph.D. degree with the School of Information and Communication Engineering, University of Electronic Science and Technology of China (UESTC), Chengdu, China.

His research interests include reconfigurable intelligent surfaces, flexible intelligent metasurfaces, and array signal processing.
\end{IEEEbiography}

\begin{IEEEbiography}
[{\includegraphics[width=1in,clip,keepaspectratio]{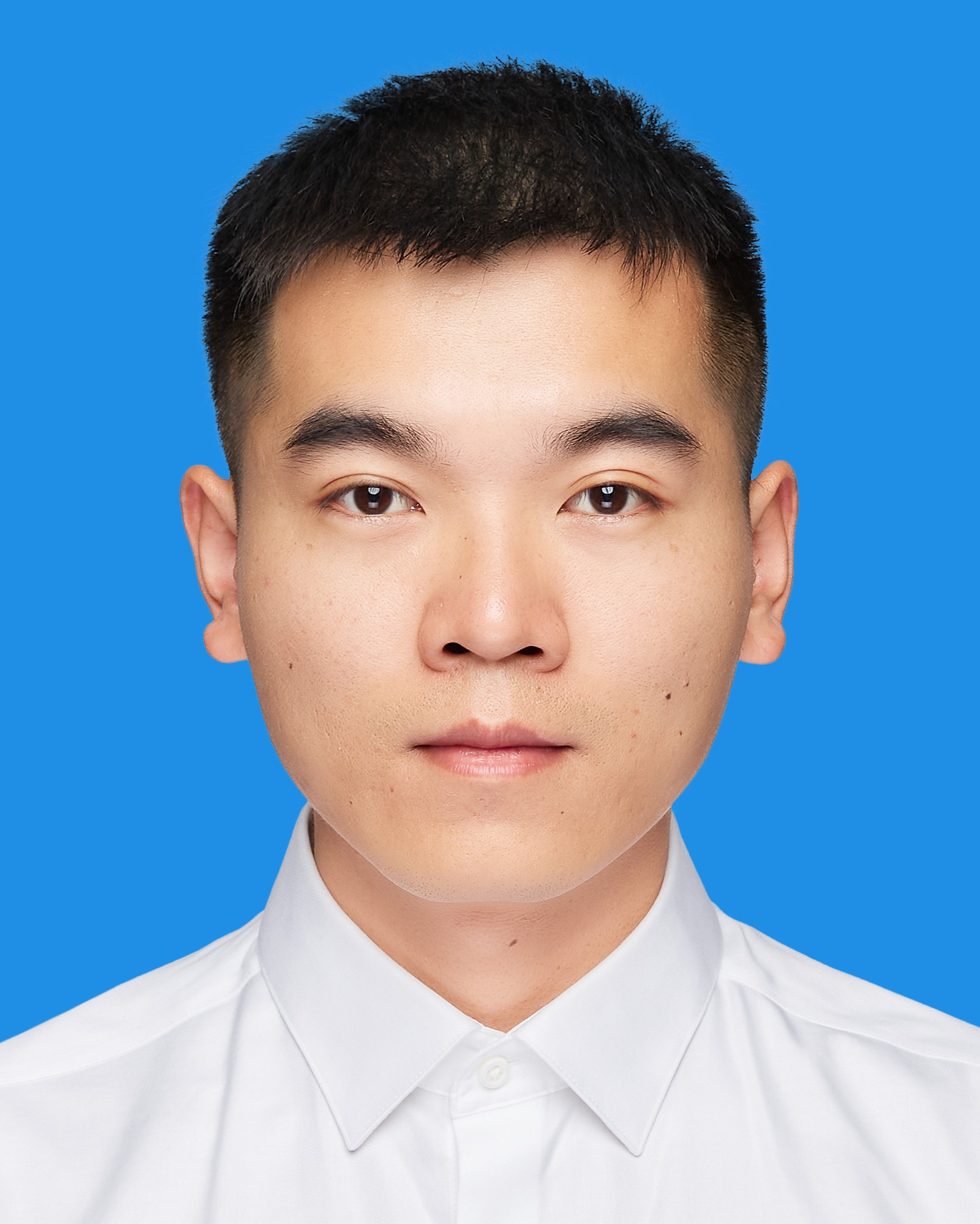}}]
{Jiancheng An} (Senior Member, IEEE) received the B.S. degree in Electronics and Information Engineering and the Ph.D. degree in Information and Communication Engineering from the University of Electronic Science and Technology of China (UESTC), Chengdu, China, in 2016 and 2021, respectively. From 2019 to 2020, he was a Visiting Scholar with the Next-Generation Wireless Group, University of Southampton, Southampton, U.K. From 2021 to 2023, he was a Post-Doctoral Research Fellow with the Engineering Product Development (EPD) Pillar, Singapore University of Technology and Design (SUTD), Singapore. From 2023 to 2026, he was a Research Fellow with the School of Electrical and Electronics Engineering, Nanyang Technological University (NTU), Singapore. He is currently a Professor with the School of Electronic Science and Engineering, UESTC, Chengdu, China.
Dr. An received the IEEE International Conference on Communications (ICC) 2023 Best Paper Award. He is the co-inventor of six patents and has published over 100 research papers in peer-reviewed international journals and conferences. His research interests include stacked intelligent metasurfaces (SIM), flexible intelligent metasurfaces (FIM), and electromagnetic neural networks (EMNN). Dr. An serves as an Editor for IEEE Transactions on Communications, IEEE Open Journal of the Communications Society, and IEEE Wireless Communications Letters. He is also the Lead Guest Editor for the Special Issue on “Stacked Intelligent Metasurface-Empowered Advanced Signal Processing Paradigm for 6G and Beyond” in IEEE Wireless Communications.

His research interests include reconfigurable intelligent surfaces, flexible intelligent metasurfaces, and array signal processing.
\end{IEEEbiography}

\begin{IEEEbiography}
[{\includegraphics[width=1in,clip,keepaspectratio]{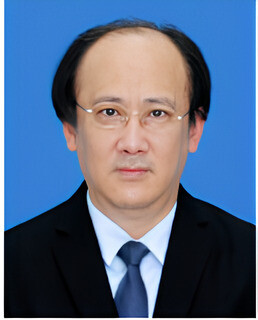}}]
{Lu Gan} (Member, IEEE) received the M.S. and Ph.D. degrees from the University of Electronic Science and Technology of China (UESTC), Chengdu, China, in 2002 and 2006, respectively.

From September 2012 to September 2013, he was a Visiting Researcher with the University of Concordia, Montreal, QC, Canada. Since August 2014, he has been a Professor with UESTC. His research interests include signal detection and classification, array signal processing, compressive sensing, passive radar, and reconfigurable intelligent surfaces.
\end{IEEEbiography}

\begin{IEEEbiography}[{\includegraphics[width=1in,clip,keepaspectratio]{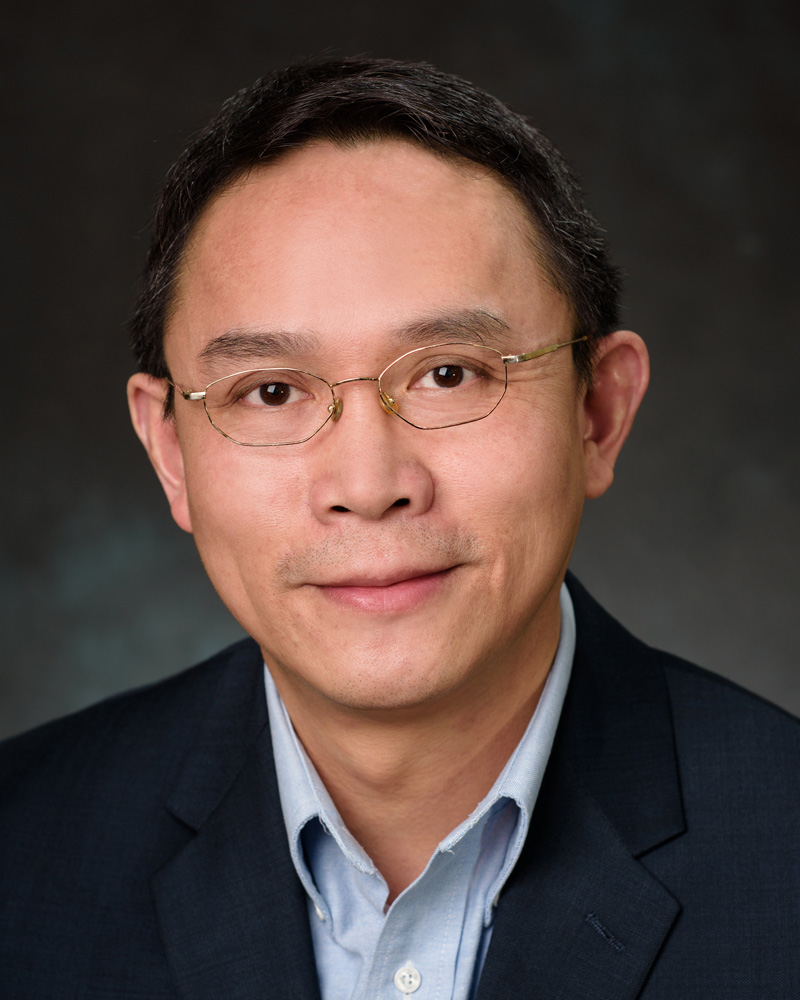}}] {Hongbin Li} (Fellow IEEE) received the B.S.~and M.S.~degrees from the University of Electronic Science and Technology of China, in 1991 and 1994, respectively, and the Ph.D. degree from the University of Florida, Gainesville, FL, in 1999, all in electrical engineering.

From July 1996 to May 1999, he was a Research Assistant with the Department of Electrical and Computer Engineering at the University of Florida. Since July 1999, he has been with the Department of Electrical and Computer Engineering, Stevens Institute of Technology, Hoboken, NJ, where he is currently the Charles and Rosanna Batchelor Memorial Chair Professor. He was a Summer Visiting Faculty Member at the Air Force Research Laboratory in the summers of 2003, 2004 and 2009. His general research interests include statistical signal processing, wireless communications, and radars.

Dr. Li received a number of awards including the IEEE Jack Neubauer Memorial Award in 2025 and 2013, IEEE Signal Processing Letters Best Paper Award in 2024, Master of Engineering (Honoris Causa) from Stevens Institute of Technology in 2024, Provost's Award for Research Excellence in 2019, Harvey N. Davis Teaching Award in 2003, and Jess H. Davis Memorial Research Award in 2001, and Sigma Xi Graduate Research Award in 1999. He was a member of the IEEE SPS Signal Processing Theory and Methods Technical Committee (TC) and the IEEE SPS Sensor Array and Multichannel TC,  an Associate Editor for \emph{Signal Processing} (Elsevier), \emph{IEEE Transactions on Signal Processing}, \emph{IEEE Signal Processing Letters}, and \emph{IEEE Transactions on Wireless Communications}, as well as a Guest Editor for \emph{IEEE Journal of Selected Topics in Signal Processing} and \emph{EURASIP Journal on Applied Signal Processing}. He has been involved in various conference organization activities, including serving as a General Co-Chair for the 7th IEEE Sensor Array and Multichannel Signal Processing (SAM) Workshop, Hoboken, NJ, June 17-20, 2012. Dr.~Li is a member of Tau Beta Pi and Phi Kappa Phi, and a fellow of the Asia-Pacific Artificial Intelligence Association (AAIA) and the International Artificial Intelligence Industry Alliance (AIIA).
\end{IEEEbiography}

\begin{IEEEbiography}
[{\includegraphics[width=1.25in,height=1.35in,clip,keepaspectratio]{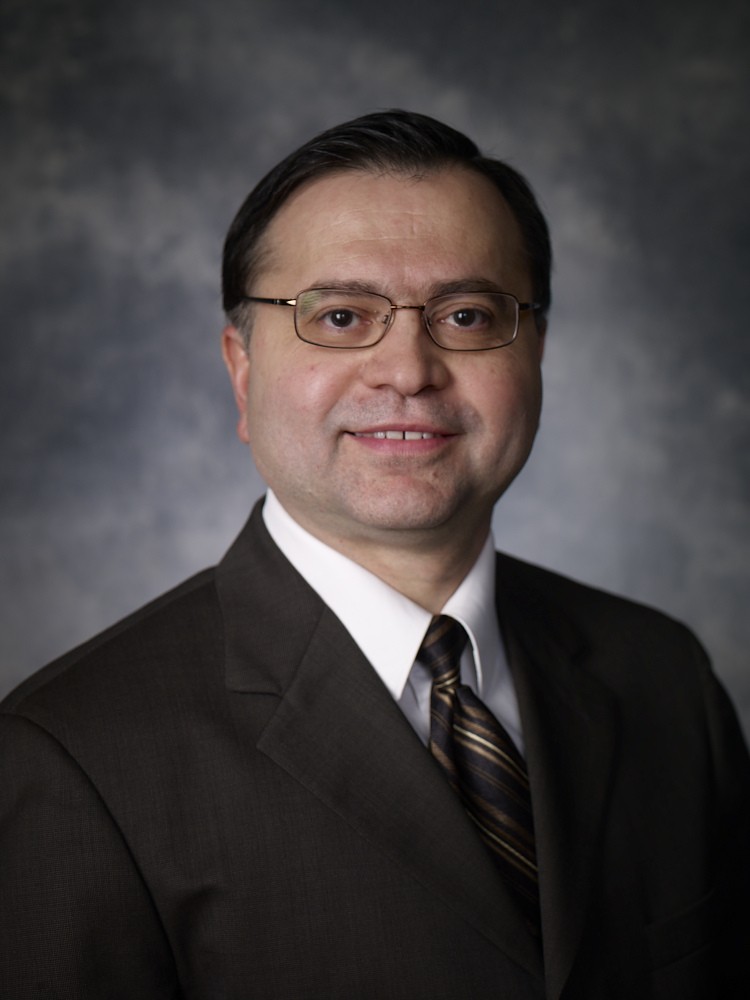}}] {Naofal Al-Dhahir} (Fellow IEEE) is Erik Jonsson Distinguished Professor and ECE Associate Head at UT-Dallas. He earned his PhD degree from Stanford University and was a principal member of technical staff at GE Research Center and AT\&T Shannon Laboratory from 1994 to 2003.  He is co-inventor of 43 issued patents, co-author of over 680 papers and co-recipient of 9 IEEE best paper awards. He is an IEEE Fellow, AAIA Fellow, received 2019 IEEE COMSOC SPCC technical recognition award, 2021 Qualcomm faculty award, and 2022 IEEE COMSOC RCC technical recognition award. He served as Editor-in-Chief of IEEE Transactions on Communications from Jan. 2016 to Dec. 2019.  He is a Fellow of the US National Academy of Inventors, a Member of the European Academy of Sciences and Arts, and a Web of Science Clarivate Highly Cited Researcher.
\end{IEEEbiography}

\begin{IEEEbiography}
[{\includegraphics[width=1.25in,height=1.35in,clip,keepaspectratio]{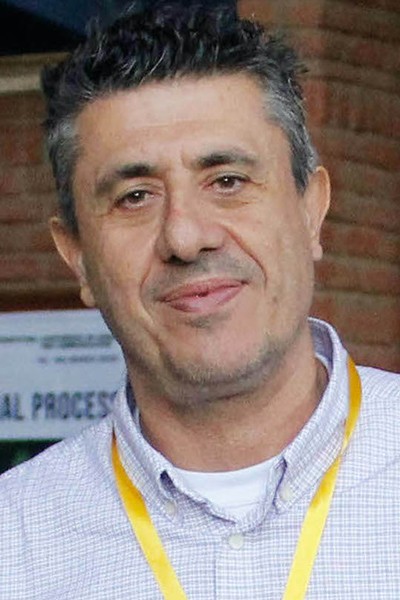}}] {George K. Karagiannidis} (Fellow, IEEE) received the Ph.D. degree in telecommunications engineering from the Electrical Engineering Department, University of Patras, Greece, in 1998. He is currently a Professor with the Electrical and Computer Engineering Department, Aristotle University of Thessaloniki, Thessaloniki, Greece, and the Head of the Wireless Communications and Information Processing (WCIP) Group. His research interests include wireless communications systems and networks, signal processing, and optical wireless communications. He has received three prestigious awards: the 2021 IEEE ComSoc RCC Technical Recognition Award, the 2018 IEEE ComSoc SPCE Technical Recognition Award, and the 2022 Humboldt Research Award from Alexander von Humboldt Foundation. He is one of the Highly Cited Authors across all areas of electrical engineering, recognized from Clarivate Analytics as the Web-of-Science Highly-Cited Researcher for ten consecutive years 2015–2024. He is also the Editor-in-Chief of IEEE Transactions on Communications and in the past was the Editor-in-Chief of IEEE Communications Letters.
\end{IEEEbiography}

\begin{IEEEbiography}
[{\includegraphics[width=1.25in,height=1.35in,clip,keepaspectratio]{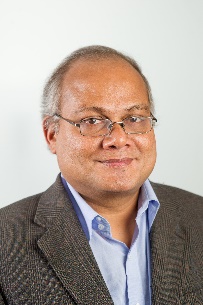}}] {Arumugam Nallanathan} (Fellow IEEE) (S'97-M'00-SM'05-F'17) is Professor of Wireless Communications and the Founding Head of the Communication Systems Research (CSR) group in the School of Electronic Engineering and Computer Science at Queen Mary University of London since September 2017. He was with the Department of Informatics at King’s College London from December 2007 to August 2017, where he was Professor of Wireless Communications from April 2013 to August 2017 and a Visiting Professor from September 2017 till August 2020. He was an Assistant Professor in the Department of Electrical and Computer Engineering, National University of Singapore from August 2000 to December 2007. His research interests include Artificial Intelligence for Wireless Systems, Beyond 5G Wireless Networks and Internet of Things (IoT). He is a co-recipient of the Best Paper Awards presented at the IEEE International Conference on Communications 2016 (ICC'2016), IEEE Global Communications Conference 2017 (GLOBECOM'2017) and IEEE Vehicular Technology Conference 2018 (VTC'2018). He is also a co-recipient of IEEE Communications Society Leonard G. Abraham Prize in 2022. He is an IEEE Distinguished Lecturer. He has been selected as a Web of Science Highly Cited Researcher in 2016, and 2022-2025. 

He was a Senior Editor for IEEE Wireless Communications Letters, an Editor for IEEE Transactions on Wireless Communications, IEEE Transactions on Communications, IEEE Transactions on Vehicular Technology and IEEE Signal Processing Letters. He served as a Guest Editor for numerous special issues of IEEE Journal on Selected Areas in Communications (JSAC). He served as the Chair for the Signal Processing and Communication Electronics (SPCE) Technical Committee of IEEE Communications Society and Technical Program Chair and member of Technical Program Committees in numerous IEEE conferences. He received the IEEE Communications Society SPCE outstanding service award 2012 and IEEE Communications Society RCC outstanding service award 2014.
\end{IEEEbiography}

\end{document}